\documentstyle[11pt]{article}

\begin{document}
\begin{titlepage}
\begin{center}
June, 1995      \hfill     HUTP-95/A021 \\
\vskip 0.2 in
{\large \bf A LATTICE CONSTRUCTION \\OF CHIRAL GAUGE THEORIES}\footnotetext{~}

\vskip .2 in
       {\bf  Pilar Hern\'andez\footnote{hernandez@huhepl.harvard.edu.
Junior Fellow, Harvard Society of
Fellows. Supported by the Milton Fund of Harvard University and by
 the National Science Foundation under grant NSF-PHYS-92-18167}} \\
and \\
         {\bf   Raman Sundrum\footnote{sundrum@huhepl.harvard.edu.
Supported by NSF under grant NSF-PHYS-92-18167}}
        \vskip 0.3 cm
       {\it Lyman Laboratory of Physics \\
Harvard University \\
Cambridge, MA 02138, USA}
 \vskip 0.7 cm

\begin{abstract}

We formulate chiral gauge theories non-perturbatively, using
two different cuttoffs for the fermions and gauge bosons.
We use a lattice with
spacing $b$ to regulate the gauge fields in standard fashion, while
computing the chiral fermion determinant on a finer lattice with spacing $f
\ll b$. This determinant is computed in the background of $f$-lattice
gauge fields, obtained by gauge-covariantly interpolating $b$-lattice gauge
fields. The
notorious doublers that plague lattice theories containing fermions
 are decoupled by the addition of a Wilson term. In chiral theories such a
term breaks gauge invariance explicitly. However, the advantage of the
two-cutoff regulator is that gauge invariance can be restored to
$O(f^2/b^2)$ by a {\it one-loop} subtraction
of calculable local gauge field counterterms. We show that the only
obstruction to this procedure is the presence
of an uncancelled gauge anomaly among the fermion representations.
We conclude that for practical purposes, it suffices to choose $f/b \sim
b/L$, where $L^4$ is the physical volume of the system. In our construction
it is  simple to prove the Adler-Bardeen theorem for
anomalies in global currents to all orders. The related subject of
 fermion number
violation is also studied.
 Finally, we discuss the prospects for improving the efficiency of our
algorithm.
\end{abstract}
\end{center}

\end{titlepage}

\section{Introduction}
Particle theories which go beyond the Standard Model are likely to
depend on the non-perturbative dynamics of chiral gauge theories.
It is well appreciated that an important step for theoretical and
computer investigations of these theories is their formulation on the
lattice.
The major obstacle to accomplishing this goal is the seeming
difficulty of avoiding `fermion doubling', while at the same
time maintaining chiral gauge invariance \cite{rev}.
In this paper we describe a non-perturbative lattice construction,
which  we  show leads in the continuum limit to Green functions
satisfying the Ward identities of a given chiral gauge theory, precisely
when gauge anomalies cancel among the fermion representations.

Several of the tools we employ exist in some form in the earlier literature
on the subject. The most important of these is the idea of coupling
continuum fermions to interpolated
lattice gauge fields as a step towards maintaining chiral
symmetry. Reference \cite{early} contains early work on this topic.
More recently it
has been used in the context of maintaining ungauged chiral
symmetry by 't Hooft \cite{thooft}. Also,
the authors of \cite{schier} and very recently \cite{hsu} and \cite{kron}
 have used a gauge field interpolation to
the continuum in defining a fermion determinant
for chiral gauge theories.

Our chiral gauge theory
construction makes use of two lattices: the gauge fields live on the
links of a regular lattice with spacing $b$, while the fermions live on a
much finer lattice subdividing the $b$-lattice, with spacing $f \ll b$.
The gauge field action is the standard one.
For the fermion-gauge boson interaction we gauge-covariantly interpolate the
$b$-lattice gauge fields to $f$-lattice gauge fields and couple these
to the fermions. If the interpolation is `smooth' enough
this will result in a regulated theory in which the gauge boson momenta are
 cutoff above $1/b$, while the fermion momenta are at most of order $1/f$.
The utility of having these two cutoffs is easy to understand
on very general grounds and we outline these next.

The Nielsen-Ninomiya  theorem \cite{nielsen} strongly suggests that in order to
remove unwanted doublers  from the spectrum of light fermion modes on the
lattice,
the chiral gauge symmetry must be broken explicitly. Resigning ourselves
to this conclusion,
there is a simple lattice fermion regulator which eliminates the doubler-modes
with a Wilson term
and results in a gauge non-invariant fermion determinant.
However, this breakage of gauge symmetry can be cured by renormalization. The
 approach of
restoring chiral symmetries through the addition of counterterms is well-known
to work in lattice QCD \cite{smit}. The case of chiral gauge theories
has been considered
 by the Rome group \cite{rome} \cite{rome2}, and also by the
authors of \cite{zaragoza}.
They propose to recover a
gauge invariant theory by a non-perturbative tuning of counterterms.
The major advantage of our approach is  that the
use of two cutoffs allows us to recover a symmetric continuum limit
after an  {\it analytic one-loop subtraction}.

Due to the Wilson term, the fermionic one-loop effective action
is not gauge invariant. Under a gauge transformation $\delta^a_x$
on the lattice, it transforms as,
\begin{equation}
\delta^a_x \Gamma_0  = L^a(x)  + O(f^2),
\end{equation}
where $L^a(x)$ is a sum of {\it local}, dimension $\leq 4$ operators made
from the gauge fields,
 while the $O(f^2)$ terms may be non-local in the gauge
fields. This form is not surprising. Any sensible regulator with
short-distance cutoff $f$ should
give  gauge invariant results in the $f \rightarrow 0$ limit, except in the
presence of ultraviolet divergences, which are sensitive to the
details of the cutoff procedure. The local nature of these ultraviolet
divergences underlies the locality of $L^a$. Any other violation of
 gauge symmetry must vanish as $f \rightarrow 0$.

The fact that the large violations of gauge invariance are local suggests
that they can be subtracted by defining a `renormalized' fermion
effective action,
\begin{eqnarray}
\Gamma_r = \Gamma_0 - S_{c.t.}, \nonumber \\
\delta^a_x S_{c.t.} = L^a(x),
\end{eqnarray}
where $S_{c.t.}$ is local in the gauge fields. The reason that
the locality of the subtractions is important is that the non-local
`finite' parts represent the consequences of one-loop unitarity and must be
left intact. Clearly $\Gamma_r$ is now gauge invariant up to order $f^2$.
This procedure provides a nice understanding of chiral gauge anomalies.
The (consistent) anomaly \cite{zumino} is the only term in $L^a$
 which is not the gauge
variation of a {\it local} functional. Therefore it cannot be
subtracted. Gauge invariance up to $O(f^2)$ can only be restored
in theories for which the gauge anomaly cancels among the chiral fermion
representations. Obviously, we will assume this to be the case.

If we had exactly cancelled all gauge non-invariance we could proceed to
integrate over gauge fields without further anxiety, since the gauge
fields are gauge-invariantly regulated on the lattice.
But this is not the case, since there remains an $O(f^2)$ non-local
violation of gauge invariance. When the fermion determinant is inserted in
gauge boson loops, new divergences threaten to overcome the $O(f^2)$
suppression. This is where having two cutoffs
makes a crucial difference. The point is that the new gauge boson
ultraviolet divergences are at most given by powers of $1/b \ll 1/f$.
Therefore the violations of gauge invariance are suppressed by the small
parameter $f^2/b^2$ to all orders! This is the central mechanism in our
construction for reproducing a gauge invariant theory in the continuum limit.

A regulator with two cutoffs, $f \ll b$, may seem strange at first,
but it is simple to understand if we imagine integrating down
 to a theory with a single
cutoff, $b$. One could in principle integrate out fermion loop momenta
 between the scales
$1/f$ and $1/b$, to yield a theory cut off at $b$, but which contains a
series of higher dimensional local interactions resulting from this
integration.
The new theory is equivalent to the two-cutoff theory for amplitudes with
external momenta below $1/b$. In particular, though gauge invariance is not
manifest, the higher-dimensional interactions will  compensate the
gauge non-invariance of the cutoff procedure to yield amplitudes related by
the gauge Ward identities to order $f^2/b^2$.

The remainder of this paper is organized as follows.
In section 2, we begin with the formal path integral for a chiral gauge
theory and explain how the various parts are to be formulated on two
lattices, focussing mostly on the fermions. In particular we explain how
the loss of gauge invariance in the regulated fermion determinant can be
compensated by subtraction of local counterterms.
The obstruction provided by
gauge anomalies to doing this is emphasized.
We also propose a hybrid of this fermion regulator
and the standard Wilson regulator for vector-like theories. It
separately regulates the real and imaginary parts of the fermion effective
action, and enjoys improved gauge invariance properties over the original
regulator, which will be important in section 4. Section 3 describes our
adaptation of 't Hooft's gauge field interpolation scheme, stressing the
 important properties it satisfies. In section 4, we study the
scaling  of the effective action with $f$,
before integration over gauge fields, taking into account the $f$ dependence
introduced by the interpolation. We show that the effective action in the
 hybrid scheme is gauge
invariant to $O(f^2)$  and finite in the limit $f\rightarrow 0$.
In section 5, using the results of section 4, it is shown that the gauge Ward
identities (or BRST identities if gauge-fixing is employed) hold to
$O(f^2/b^2)$ after functional integration over gauge fields.
The choice of how small to take $f/b$ in practice is then made.
Section 6 deals with
anomalies in global currents which occur in chiral theories. The
Adler-Bardeen theorem is demonstrated to hold at one loop and then  proven to
hold to all orders. The second step is made particularly simple by the
use of the two-cutoff regulator. Section 7 outlines how the anomalies in
global currents translate into the phenomenon of fermion number violation
in our construction. Section 8 discusses some new ideas for
improving computational efficiency. Section 9 provides our conclusions.

\section{Two-Cutoff Lattice Chiral Gauge Theory}

Our target theory is one describing a left-handed fermion with a
non-abelian gauge charge.
The continuum version which we want to regulate with the lattice is formally
 given by the generating functional
\begin{eqnarray}
e^{W[J_{\mu}, \overline{\eta}, \eta]}& =& \int {\cal D} A_{\mu} {\cal D}
\overline{\psi} {\cal D} \psi \; \exp^{- \int \frac{1}{4}tr
F_{\mu \nu}^2 + \overline{\psi}
i \hat{D} \psi + J_{\mu} A_{\mu} + \overline{\eta} \psi + \overline{\psi}
\eta} \nonumber \\
& = & \int {\cal D} A_{\mu}\; e^{- \int \frac{1}{4}
tr F_{\mu \nu}^2 + J_{\mu} A_{\mu}}~ e^{\Gamma[A]}  e^{- \int \int
\overline{\eta} G[A] \eta},
\label{target}
\end{eqnarray}
where
\begin{eqnarray}
e^{\Gamma[A]} &=& {\rm det}\; i \hat{D}, \nonumber \\
G[A] &=& (i \hat{D})^{-1}, \nonumber \\
\hat{D} &=& (\not\!\partial + i \not\!\!A)\; L~ + \not\!\partial\; R,
\end{eqnarray}
and $R, L = \frac{1}{2}(1 \pm \gamma_5)$. We see that $\psi_L$ transforms
(in general reducibly) under the gauge
group, while $\psi_R$ is taken to be sterile.

The above manipulations are valid assuming that  $i\hat{D}$ has no
zero-modes, which would leave $G[A]$ ill-defined. However, the
occurrence of zero modes in topologically non-trivial gauge field
backgrounds is physically important, for example in fermion-number
violating processes. For most of this paper we will restrict our
attention to fermion number conserving processes. In sections 6 and 7
we show that fermion number violation occurs in our construction and we
describe a procedure for defining general fermion number violating amplitudes.

Our lattice construction of the target theory (\ref{target}) will take the
form
\begin{equation}
e^{W[J_{\mu}, \overline{\eta}, \eta]}
 = \int \prod_s d U_{\mu}(s)~ e^{- S_{g}[U] + \sum_s J_{\mu}
A_{\mu}} ~e^{\Gamma[U]} ~ e^{- \sum
\overline{\eta} G[U] \eta},
\label{effact}
\end{equation}
where $U_{\mu}(s) = e^{i b A_{\mu}(s)}$ is the gauge field variable on the
link $(s, s + \hat{\mu})$ of a regular lattice with spacing $b$.
The gauge boson kinetic terms are defined in the usual way on the
$b$-lattice,
\begin{eqnarray}
S_{g}[U] = \frac{2}{g_0^2} \sum_s \sum_{\mu<\nu}
 [I - \frac{1}{2}(U_{\mu\nu}+ U_{\mu\nu}^\dagger)],
\nonumber\\
U_{\mu\nu} \equiv U_\mu(s)
U_\nu(s+\hat{\mu})
U_\mu^{-1}(s+\hat{\nu}) U_\nu^{-1}(s).
\label{gauact}
\end{eqnarray}

To define the fermion determinant however, we employ a much finer lattice
which subdivides the $b$-lattice, with
spacing $f \ll b$, where $b/f$ is an integer. The fermion determinant is then
given by the $f$-lattice path integral  with action
\begin{eqnarray}
S^{\chi w}_f[\Psi] = \frac{f^3}{2} \sum_{n,\mu} (\bar{\Psi}(n) \gamma_\mu
[R + u_\mu L]
\Psi(n+\hat{\mu})
 - \bar{\Psi}(n+\hat{\mu}) \gamma_\mu [R + u^{-1}_\mu L] \Psi(n)\nonumber\\
+ 4 r f^3\sum_n \bar{\Psi}(n) \Psi(n) - \frac{r  f^3}{2} \sum_{n,\mu}
[\bar{\Psi}(n) \Psi(n+\hat{\mu})
+\bar{\Psi}(n+\hat{\mu}) \Psi(n)]\},
\label{cw}
\end{eqnarray}
where $n$ denotes the points of the
$f$-lattice (as opposed to $s$ for the points of the $b$-lattice).
We refer to this as the chiral
Wilson ($\chi w$) lattice regulator to distinguish it from the standard Wilson
($sw$) regulator for vector-like gauge theories \cite{rothe}. It is also the
regulator used by the Rome group \cite{rome} for $f = b$.
The action describes a left-handed fermion coupled to
a gauge field $u_{\mu}$, and a sterile right-handed fermion. The two
chiralities are
coupled together by a Wilson term proportional to $r$, which gives the
left- and right-handed doubler modes a Dirac mass of order $r/f$
 but clearly breaks
chiral gauge symmetry explicitly.
We take $r =O(1)$. The explicit breaking of gauge invariance is a
nuisance but not fatal. As we will show, it can be made small {\it to all
orders}  by the addition of gauge field counterterms determined at one
(fermion) loop.
Note that we have not added a mass for the fermion. It is not required
 because the
$\chi w$ action possesses a `shift' symmetry, $\psi_R \rightarrow \psi_R +
\xi$,
where $\xi$ is a constant spinor \cite{golt}. On the other hand, this
 is not a symmetry of
the standard Wilson term of vector-like
theories and, as is well-known, mass counterterms are required in that
context\footnote{In any case, in the two-cutoff
theory only a mass of $O(f/b^2)$ would be generated.}.

 The $f$-lattice link variables, $u_\mu(n)$, are interpolations of the
$b$-lattice gauge fields, $U_\nu(s)$,
which are the real dynamical variables. The one-loop fermion
effective action (the logarithm of the fermion determinant)
 will therefore be a functional of the $b$-lattice gauge fields,
\begin{equation}
\Gamma[U] = \Gamma[u[U]].
\end{equation}
 The interpolation $u[U]$ must be
carefully chosen so that a (nearly) gauge invariant, lattice rotation and
translation invariant, determinant functional
of $u_{\mu}$ induces a (nearly) gauge invariant, lattice rotation and
translation invariant  functional of $U_{\mu}$. Furthermore, the
non-locality of $\Gamma[u]$ correctly describes the effects of virtual
massless fermions dictated by one-loop unitarity, and it is essential that
this same non-locality be found in the $\Gamma[U]$ induced by the
interpolation. In section 3, we will describe
such an interpolation and its properties in detail.
For the remainder of this section we will assume that we
have such an interpolation procedure and concentrate on compensating the
breakage of gauge invariance (in terms of $u_{\mu}$) inherent to the $\chi w$
regulator.

We will denote the unsubtracted (gauge non-invariant) $\chi w$ determinant by
$\exp \Gamma_0[u]$.
Consider the effect of an infinitesimal gauge transformation,
\begin{equation}
u_{\mu}(n) \rightarrow \omega(n) u_{\mu}(n) \omega^{-1}(n + \hat{\mu}),
\end{equation}
with $\omega(n) \equiv e^{i \theta^a(n) T^a}$.
 We have then,
\begin{eqnarray}
\delta_{\omega} \Gamma_0[u] &=& <\delta_{\omega} S^{\chi w}_f>_u, \nonumber \\
\delta_{\omega} S^{\chi w}_f &=& - i 4  f^3 \sum_n ~\theta^a(n)
\{ \bar{\Psi}(n)~ \gamma_5~  T^a \Psi(n)\nonumber\\
&+& \frac{1}{8} \sum_\mu [(\bar{\Psi}(n-\hat{\mu})+
\bar{\Psi}(n+\hat{\mu}))~L~ T^a \Psi(n) + h.c. ] \}
\label{omega}
\end{eqnarray}
where only the Wilson terms have contributed to the non-invariance of the
action, and $<..>_u$ denotes the fermionic expectation in the background of
the gauge field.

Naively the breakage of gauge invariance is only of order
$f^2$. This is because the Wilson term corresponds to a
dimension-five interaction suppressed by $f$ for the physical modes, and to a
mass of $O(1/f)$ for the doubler modes, whose effects one might expect to
be suppressed by their mass. Furthermore, these interactions are chirality
flipping and must  therefore come in pairs. This naive
expectation is however upset by power-like ultraviolet divergences in $1/f$ in
the one loop computation of (\ref{omega}), which can cancel the $f^2$
suppression. Of course these ultraviolet divergent parts are local
functionals of the gauge fields and $\omega$, the allowed forms of which
can be determined by power-counting, lattice rotational symmetry and
invariance under space-time independent gauge transformations.
 The finite parts of $\delta_{\omega} \Gamma_0[u]$ can be non-local in
$u_{\mu}$, but they are suppressed by the power of $f^2$ just discussed.
Thus,
\begin{equation}
\delta_{\omega} \Gamma_0[u] = \sum_n L(n) + O(f^2),
\end{equation}
where $L(n)$ is a local function of $\omega$ and $u_{\mu}$. $L(n)$ can contain
local operators up to dimension 5.

We are now in a position to see the difference between the breakage of
gauge invariance caused by a non-invariant regulator and that caused by
having an uncancelled gauge anomaly. For example, one of the terms allowed
in $L(n)$ is
\begin{equation}
L_1(n) = \frac{\alpha}{f^4} \sum_{\mu} {\rm Tr}[\omega(n) u_{\mu}(n)
\omega^{-1}(n+\hat{\mu}) - u_{\mu}(n)] + h.c.
\end{equation}
 It is invariant under lattice rotations and it vanishes for
constant $\omega$. The power of $1/f$ is dictated by dimensional
analysis since $f$ is the only scale in $\Gamma_0$. The coefficient $\alpha$
however, must be extracted from a one-loop lattice diagram. Although the
breakage of gauge invariance given by this term is large, the fact that it
is local means we can substract it from $\Gamma_0$ if there exist a local
 functional of $u_\mu$,
$S_{c.t.}[u]$ such that $\delta_{\omega} S_{c.t.} = L_1$.
For the present example such a
subtraction can be performed,
\begin{eqnarray}
\Gamma_r[u] &=& \Gamma_0 - S^1_{c.t.}[u], \nonumber \\
S^1_{c.t.}[u] &=& \frac{\alpha}{f^4} \sum_{\mu}
{\rm tr}(u_{\mu}(n) - 1)(u^{-1}_{\mu}(n) - 1),
\end{eqnarray}
which contains a gauge boson mass counterterm.
We are only allowed to subtract
local functionals because the non-local part of $\Gamma_0$ already
correctly reflects the requirements of one-loop unitarity. If a term in
$L(n)$ can be cancelled by subtraction of a local functional up to  $O(f^2)$
we will refer to it as `locally integrable'.

In the absence of gauge anomalies, all the possible terms in $L(n)$ are
 locally integrable. This result
for the $\chi w$ action was first proved in \cite{rome} at one loop.
The full set of gauge field counterterms are\footnote{We follow the
notation of ref. \cite{rome}. We however do not
include the operators involving ghosts since they are not needed in our
two-cutoff  construction.},
\begin{itemize}
\item Mass Counterterms:
\begin{eqnarray}
\mu_W^2 \frac{1}{f^2}\; Tr[\sum_\mu (u_\mu(n)-1)(u_{\mu}^{-1}(n) -1)]
\end{eqnarray}
\item Wave Function Renormalization:
\begin{eqnarray}
\frac{1}{f^4}[ \delta Z_W Tr\;\sum_{\mu\nu} ( \Delta_\nu u_\mu )^2
+ \delta \zeta Tr\;(\sum_{\mu} \Delta_\mu u_\mu )^2
+ \delta Z_{NC} Tr\;\sum_{\mu} ( \Delta_\mu u_\mu )^2]
\end{eqnarray}
\item Gauge Boson Self Interaction:
\begin{eqnarray}
\delta Z_3  \frac{1}{f^4} Tr \sum_{\mu,\nu} \Delta_\mu u_\nu [u_\mu,
u_\nu],
\nonumber\\
\delta Z_{4,A}  \frac{1}{f^4} Tr \sum_\mu (u_\mu -1)^2 Tr
\sum_\mu (u_\mu-1)^2,\nonumber\\
\delta Z_{4,B}   \frac{1}{f^4} \sum_{\mu,\nu} Tr[ (u_\mu-1)(u_\nu-1)]
Tr[(u_\mu-1)(u_\nu-1)],
\nonumber\\
\delta Z_{4,NC}   \frac{1}{f^4} \sum_\mu Tr[(u_\mu -1)^2 \;(u_\mu -1)^2].
\end{eqnarray}
\end{itemize}
The $\Delta_{\mu}$ are lattice derivatives.
So, if $L(n)$ is locally integrable, a one-loop subtraction yields
 $\delta_{\omega} \Gamma_r[u] = O(f^2)$ and, despite the lack of a gauge
invariant regulator, gauge invariance is recovered in the continuum limit $f
\rightarrow 0$, at least at one loop.

The only term that can occur in $L(n)$ which is not
locally integrable is the consistent anomaly,
\begin{equation}
L^a(n) = -\frac{i}{12 \pi^2}
{\rm Tr}~ \theta(n)~ \epsilon_{\alpha \beta \gamma \delta}
{}~\Delta_{\alpha} [a_{\beta} \Delta_{\gamma} a_{\delta} - \frac{i}{2}
a_{\beta} a_{\gamma} a_{\delta}] + O(f^2),
\end{equation}
where $u_{\mu}(n) = e^{i f a_{\mu}(n)}, \omega(n) = e^{i \theta(n)}$.
A brief description of the derivation of this expression on the lattice is
given
in section 6. The gauge color trace is
proportional to the anomaly coefficient, $d_{abc} \equiv \frac{1}{2}
 Tr[T^a \{T^b, T^c\}]$, of the fermion
representations. If it is not zero, the
anomaly breaks gauge invariance and cannot be subtracted away with any local
counterterms. We will assume that  all gauge anomalies cancel. Hence
$\Gamma_r[u]$ is gauge invariant up to $O(f^2)$.

Having dealt with gauge invariance (at one loop), let us consider the
physics that the fermion effective action incorporates.
We have used the freedom
one has in regulating fields in the ultraviolet to impose two
cutoffs on the theory. The physics at distances $\gg b \gg f$ which survive the
continuum limit will be insensitive to whether we use one or two cutoffs. In
principle we could first integrate out all fermion momenta above
$1/b$ and get a local theory with just one momentum cutoff. It would contain
extra higher dimension interactions dependent on $b$ and $f$, some of which
would break gauge invariance. But this breakage would be cancelled by the
non-invariance of the cutoff procedure to yield just an $O(f^2)$ breaking of
gauge invariance in the one-loop amplitudes of the theory.

This brings us
to an important point. We have just seen that all unsuppressed gauge
non-invariance at one loop can be subtracted away (when gauge anomalies
cancel). What
remains of the gauge non-invariance is insensitive to the $f \rightarrow 0$
limit. But
there will be one gauge invariant
interaction induced by integrating out fermion momenta
between $1/f$ and $1/b$,
which will  be sensitive to $f \rightarrow 0$. This is the gauge
field kinetic term which   must
be renormalized by integrating out fermion momenta between $1/f$ and $1/b$.
The result for its coefficient is,
\begin{equation}
1/g_b^2 = 1/g_0^2 + \log(b/f) \frac{t_2(F)}{12 \pi^2},
\label{coupling}
\end{equation}
where $t_2(F)$ is the sum of quadratic Casimir invariants for the fermion
representations. Recall that $1/g_0^2$ is the original gauge coupling of the
two-cutoff theory, multiplying the gauge kinetic term, $S_g$. The second
term on the right-hand side is just the one-loop renormalization induced by
integrating out fermion momenta between the two cutoffs. Therefore $g_b$ is
to be identified as the single true coupling of the effective one-cutoff
theory which must be
scaled to zero in the continuum limit $b \rightarrow 0$. In terms of $g_b$,
the action of the effective one-cutoff theory is insensitive to $f
\rightarrow 0$ (assuming, as we have up till now, that our interpolation
procedure does not reintroduce sensitivity to $1/f$).
Having identified the correct physics within the effective
one-cutoff theory,
we can proceed to always use the explicitly constructed two-cutoff theory
for practical purposes.

In a straightforward one-cutoff theory with cutoff $1/f$,
the one-loop suppression of gauge invariance
breaking by $f^2$ would not have bought us much. The problem is that the real
suppression factor is $f^2 \times$ gauge boson momenta$^2$. Once
$\Gamma_r[u]$ is put into gauge boson loops, gauge boson momenta of order
$1/f$ could arise, cancelling the $f^2$ suppression and leading  to
large violations of  gauge invariance. This is not the case in
a two-cutoff scheme such as ours. In loops, the gauge boson momenta are of
order $1/b$ at most since they live on the $b$-lattice links, so that
deviations from gauge invariance will be suppressed by
$O(f^2/b^2)$,
so long as the interpolation procedure is smooth enough to not
reintroduce substantial $O(1/f)$ momentum components into the interpolated
gauge
field $u$ (ie. the $b$-lattice softly cuts off the $u_{\mu}$ momenta above
$1/b$). We can then choose $f/b$ small enough for any given precision and
space-time volume we wish to work with.
Unfortunately, we will show in section 4 that the interpolation procedure
of section 3 is {\it not} smooth enough for the $\chi w$ scheme outlined here
to work! We are therefore led in the next subsection to a scheme which is a
simple hybrid of our $\chi w$
scheme and the standard Wilson ($sw$) scheme for vector-like gauge theories.
We will show  that our interpolation procedure is smooth enough
for this improved hybrid scheme to work. It is also quite possible, that a
smoother interpolation scheme can be found, in which case the $\chi w$ scheme
would be feasible.

Finally, we define the fermion propagator in an external gauge field,
$G[u]$, needed in order to define Greens functions with external fermion
lines,
\begin{equation}
G[u]^{-1}_{n_1 n_2} =
\frac{\delta}{\delta \Psi(n_1)} \frac{\delta}{\delta \overline{\Psi}(n_2)}
S^{\chi w}_f[\Psi].
\label{prop}
\end{equation}
If we only consider processes without external sterile
fermions, then we must have an even number of chirality flips from the
Wilson term in the propagator. Since it is the only term
breaking gauge invariance and we are not yet integrating over gauge loops,
it is easy to see that the propagator is gauge covariant up to order $f^2$.

\subsection{Hybrid Regulator}

We describe an alternative lattice construction of the one-loop fermionic
effective action which  separately regulates the real and
imaginary parts. It will automatically be gauge invariant in the
 limit $f \rightarrow 0$, without subtractions,
 precisely when gauge anomalies cancel among the fermion
representations. This type of separation was introduced
in ref.\cite{gaume} and further discussed in the lattice context
in refs.\cite{schier}\cite{bodwin}.
Since the trick used may be unfamiliar we
end the section with a brief discussion of the legitimacy of the regulator
in preserving the unitarity property of the unregulated fermion
determinant.

The basic, well-known observation is that formally the
magnitude of the fermion determinant, $e^{{\rm Re} \Gamma}$,
is parity-even and is just the square
root of a vector-like determinant, which we know how to formulate with
{\it exact} gauge
invariance on the lattice. The phase of the chiral fermion determinant,
Im$\Gamma$, is
 parity-odd but can be regulated by means similar to those of the previous
section, though of course this leads to some breakage of gauge invariance.
The advantage here is that this breakage is small.
As we saw in the last section, it is
given by
the sum of local dimension $\leq 4$ terms in the gauge field plus non-local
terms which are however suppressed by powers of $f^2$. It is easy to see
from the possible local terms listed in the last section that the only
parity-odd term is the consistent anomaly. For anomaly-free theories
the breakage of gauge invariance is thus of order $f^2$.

To effect the separation into Re$\Gamma$ and Im$\Gamma$, we note that
formally,
\begin{equation}
i\hat{D} = [ i\hat{D}~ (i\hat{D})^{\dagger -1}~ (i\hat{D})^{\dagger}
{}~i\hat{D}]^{\frac{1}{2}}.
\end{equation}
Taking the determinant of both sides of the above equation yields
\begin{equation}
\det[i\hat{D}] = \{\det[i\hat{D}] \;\frac{1}{\det[i\hat{D}]^*}\;
\det[(i\hat{D})^{\dagger}]\; \det[i\hat{D}]\}^{\frac{1}{2}}.
\end{equation}
Now, since $(i\hat{D})^{\dagger} = i\!\!\not\!\!D R + i\!\!\not\!\partial
L$, while
$i\hat{D} = i\!\!\not\!\!D L +
 i\!\!\not\!\partial R$, the third and fourth determinant factors
 in the square root together describe a Dirac fermion with vector-like
gauge couplings, and a sterile Dirac fermion. Thus we have,
\begin{equation}
\det[i\hat{D}] = \{\frac{\det[i\hat{D}]}{\det[i\hat{D})]^*}\;
\det[i\!\not\!\!D]\; \det[i\!\not\!\partial]\}^{\frac{1}{2}}.
\end{equation}
The last, gauge field independent determinant only contributes a constant
to the fermion effective action and can be dropped. The first
ratio is a phase factor, while $\det[i\not\!\!D]$
is a real,  vector-like determinant. Its square-root can consistently be
taken as positive as long as the theory does not suffer from Witten
anomalies \cite{witten}.   The idea of this section is to
lattice-regulate these determinants separately, in a way which manifestly
respects the even and odd parities of Re$\Gamma$ and Im$\Gamma$ respectively.

We now have at our disposal a simple means of formulating our new
improved chiral fermion determinant,
\begin{equation}
\det[i\hat{D}]_{hybrid} \equiv (\frac{\det[i\hat{D}]_{\chi w}}{
\det[i\hat{D}]_{\chi w}^*})^{\frac{1}{2}}.~
(\det[i\not\!\!D]_{sw})^{\frac{1}{2}}.
\label{hs}
\end{equation}
Here, $\det[i\hat{D}]_{\chi w}$ is just the $f$-lattice chiral determinant of
the last section, while $\det[i\!\!\not\!\!\!D]_{sw}$ is  the
standard gauge-invariant Wilson determinant for vector-like fermions.
The only difference with single-lattice vector-like theories is that a mass
counterterm is still not required, because
the mass induced by breaking chiral symmetry via the standard
Wilson term is at most $O(f/b^2)$.

As explained earlier in this section,
the only breakage of gauge
invariance in $\det[i\hat{D}]_{hybrid}$ must be parity-odd, and therefore
 suppressed by $f^2$ in anomaly free models.

Let us view the above decomposition and regularization in diagrammatic terms.
The loop diagrams representing the chiral fermion effective action
 have vertices containing $L = (1 - \gamma_5)/2$. Even before
regulating the diagrams one can separate the real (parity-even) and
imaginary (parity-odd) parts of
the original diagrams. It is easy to check that the diagrams
representing (the unregulated) $\det[i\!\not\!\!\!D]^{1/2}$ and
$(\det[i\hat{D}]/\det[i\hat{D}]^*)^{1/2}$ reproduce exactly these
parts respectively. By regulating these determinants separately we are
just regulating
the parity-even and odd diagrams separately.

This scheme does not
harm the  unitarity properties possessed by the unregulated diagrams.
The point is that any diagram can be written as a non-local part
which is insensitive to the cutoff and a local part (polynomial in
external momenta) which may be sensitive
to the cutoff. Unitarity constrains only the non-local parts, the local
terms corresponding to arbitrary `subtraction constants'.

If composite operators need to be defined, they may require new counterterms
in order to recover gauge invariance. See section 6 for the example of
global currents.

Finally, amplitudes which involve external fermion lines can
continue to be treated
using the $\chi w$ propagator in the background of the gauge field.

\section{Gauge Field Interpolation}

We describe our adaptation of a recent suggestion by 't Hooft
\cite{thooft} for
interpolating a gauge field living on the
links of the $b$-lattice. We consider the two cases of interpolation to any
$f$-lattice subdivision of the $b$-lattice, or to the continuum. The latter
will be of use since many of the limiting properties of
the $f$-lattice interpolation as $f \rightarrow 0$ will be those of the
continuum interpolation. After having defined the interpolation algorithm
we discuss the important properties that it satisfies.
 An alternative interpolation which may be computationally simpler is
discussed in section 8.

Broadly speaking, the interpolation is obtained by pasting together
 solutions of the
continuum or $f$-lattice Yang-Mills equations in each four-dimensional
hypercube of the $b$-lattice, where the gauge field on the boundaries of
the hypercube are specified in the directions tangential to the boundary.
This should determine a unique gauge field in the hypercube up to gauge
transformations in its interior, which we fix with a Lorentz gauge
condition.\footnote{The existence of solutions and their uniqueness once
the Lorentz gauge condition is imposed can be proven to all orders in the
gauge field self-coupling. This result should apply when the flux through any
$b$-plaquette is sufficiently small, as will be the case near the
continuum limit.}
The tangential gauge fields on the three-dimensional boundary cubes are
themselves
determined by three-dimensional Yang-Mills equations, and similarly for their
own boundaries, the two-dimensional plaquettes. The continuum or
$f$-lattice gauge fields tangential to the links of the $b$-lattice are
fixed to be constant along the link, and these set the ultimate
boundary conditions for the multi-stage interpolation procedure.
The details now follow.

Let $U_{\mu}(s) = \exp(i b A_{\mu}(s))$ be the gauge field configuration on
the $b$-lattice. We can describe the interpolation to either the continuum or
the $f$-lattice in terms of a field $a_{\mu}(x)$, which represents the
continuum gauge field, or the $f$-lattice gauge field via
$u_{\mu}(x) = \exp(i f a_{\mu}(x))$ (in which case $x$ is taken to be
restricted to the $f$-lattice vertices).
First, on points lying on the links of the
$b$-lattice we interpolate the tangential gauge field component by,
\begin{equation}
a_{\mu}(x = s + b \lambda \hat{\mu}) =  A_{\mu}(s),
\end{equation}
where $0 \leq \lambda < 1$.
 The interpolation into a $1$-$2$-directed $b$-lattice plaquette
 is given by minimizing the two-dimensional Yang-Mills action
\begin{equation}
S = \int dx_{1} dx_{2} f_{12}^2(x),
\end{equation}
or
\begin{equation}
S = \sum_{x} {\rm Tr}[ 1 -  u_1(x) u_2(x + f \hat{1}) u^{-1}_1(x + f \hat{1} +
f
\hat{2}) u_2^{-1}(x + f \hat{2})]  + {\rm h.c.},
\end{equation}
in the gauge $\partial_{1} a_{1} + \partial_{2} a_{2} = 0$, subject to the
boundary conditions given by agreement with the $a_{1}$ and $a_{2}$ already
determined on the bordering $1$- and $2$-directed links respectively. The
$x$ being integrated (summed) over are restricted to the points inside the
$b$-lattice plaquette.
$f_{\mu \nu}$ denotes the field strength made from
$a_{\mu}$.  Note that only the
$a_{\mu}$ components tangential to the $b$-plaquette have been determined.
The generalization to
interpolating in $b$-plaquettes pointing in other directions is obvious.

The way to complete the interpolation procedure is also clear.
We interpolate to the interior of any $b$-lattice
3-cube by minimizing
the appropriate three-dimensional Yang-Mills action in three-dimensional
Lorentz gauge with the boundary condition being agreement with the
tangential gauge field on the plaquettes bordering the cube, determined
at the previous stage. This will determine the three components of the gauge
field tangential to the cube.  Finally we fully interpolate
the gauge field to every four-dimensional hypercube by minimizing the
four-dimensional Yang-Mills action in Lorentz gauge with the boundary
conditions being agreement with the tangential gauge fields in the 3-cubes
bordering the hypercube. This completes the determination of $a_{\mu}(x)$.

This interpolation procedure is a finite computation (in finite volume)
when going from a $b$- to $f$-lattice. It satisfies several important
properties which we now list and briefly discuss.

$\bullet$ Transverse continuity:
 This describes how smooth our interpolation is.
As one passes from one hypercube of the $b$-lattice into another,
the components transverse to
the direction taken (ie. tangential to the 3-dimensional boundary) vary
continuously because these components were fixed on the boundary as part of
the boundary conditions for interpolation into each of the adjacent hypercubes.
However the `longitudinal' component of the gauge field can vary
discontinuously. This is also true at lower dimensional levels: passing
between adjacent 3-cubes or adjacent plaquettes or adjacent links of the
$b$-lattice.

One may wonder what the relevance of transverse continuity is. The point is
that the interpolation procedure  does not {\it sharply} cut off
the momenta of the $a_{\mu}$ fields above $\pi/b$ even though of course the
$b$-lattice fields we are interpolating have sharply cut off momenta.
Rather, the interpolation procedure should be seen as a `soft' cutoff of
the $a_{\mu}$ momenta above roughly $1/b$. The less smooth the
interpolation, the less sharp the cutoff of $a_{\mu}$ momentum. Now suppose
$\Gamma[a]$ is a well-defined functional of continuum gauge fields
$a_{\mu}$.
We will want $\Gamma[a[U]]$ to be a finite functional of $U$.
This will fail unless $a[U]$ momentum components fall off fast enough in
the ultraviolet. (In an $f$-lattice we will not get an infinity but will
get unsubtractable dependences on $1/f$ which will preclude our taking the $f
\rightarrow 0$ limit.)
This important consideration is taken up in detail in the
next  section  where it is shown that transverse continuity is sufficient to
ensure a well-behaved fermion determinant as a functional of $U_{\mu}$.

$\bullet$ Gauge Invariance: A gauge invariant functional of the
interpolated gauge
field, $\Gamma[a]$, automatically yields a gauge invariant functional of the
$b$-lattice gauge field, $\Gamma[a[U]$.
Clearly, it is sufficient to show that
\begin{equation}
a[U^{\Omega}] = a^{\omega}[U],
\end{equation}
where $\Omega(s)$ is a $b$-lattice gauge transformation of $U_{\mu}$, and
$\omega[U,\Omega](x)$ is a continuum ($f$-lattice) gauge transformation which
interpolates $\Omega$ and {\it can depend on $U$}. We show that such an
$\omega$ exists by describing its construction in the links, then
plaquettes, cubes and hypercubes spanned by the $b$-lattice.

 For points on a link of the $b$-lattice in the $\mu$
direction we take
\begin{equation}
\omega(x = s + \lambda \hat{\mu} b) = (\Omega(s) U_{\mu}(s) \Omega^{-1}(s +
\hat{\mu} b))^{-\lambda} \Omega(s) U_{\mu}(s)^{\lambda}.
\end{equation}
This has the property that
$a^{\omega}_{\mu}(s + \lambda \hat{\mu} b)) =
A^{\Omega}(s)$, where $U_{\mu}^{\Omega} = e^{i
A_{\mu}^{\Omega}}$, which is indeed the behavior of $a[U^{\Omega}]$ on
points on the $b$-link. Inside a $1$-$2$
directed $b$-lattice plaquette we extend
$\omega$ defined on the bounding links in a continuous fashion, so that
$a^{\omega}$ is now defined in the plaquette. Since one is always free to
take a gauge field to Lorentz gauge via a gauge transformation,
$\omega$ can be chosen so that $a^{\omega}$ is in the two-dimensional
Lorentz gauge $\partial_1 a^{\omega}_1 + \partial^{\omega}_2 a_2 = 0$.
By the gauge
invariance of the two-dimensional Yang-Mills action, $a^{\omega}$ yields the
minimum action of gauge fields agreeing with the tangential components
$a^{\omega} = A^{\Omega}$ already defined on the bounding links. These are
precisely the defining properties of the interpolation $a[U^{\Omega}]$ on
the plaquette, so $a^{\omega} = a[U^{\Omega}]$ there. Again, the remainder
of the construction should be clear; $\omega$ can be extended into cubes
and hypercubes till one has a gauge transformation everywhere and
$a^{\omega}$ satisfies the defining properties of $a[U^{\Omega}]$.

$\bullet$ Locality: When the  momentum space Feynman vertex for the coupling
of the interpolation $a_{\mu}$ to the fermions is expressed in terms of
$A_{\mu}$, the result is a series of interaction vertices which are
polynomial in the $A_{\mu}$ momenta and fields, with higher dimension
interactions balanced by powers of $b$. For low
momenta $q \ll 1/b$,
\begin{equation}
a_{\mu}(q) \sim A_{\mu}(q).
\label{aq}
\end{equation}
This allows us to view the whole interpolation procedure as an
intermediate step in coupling $b$-lattice gauge fields to fermions that live
in the continuum ($f$-lattice) in such a way that the low-energy physics
far below $1/b$ is unaffected. The complexity of the procedure
arises from the need to maintain  gauge invariance.

The reason locality is satisfied by the interpolation is quite easy to see:
$a_{\rho}$ at the point $x = s + y b$ is
determined purely by $A_{\mu}(s), A_{\mu}(s + \hat{\nu} b)$, that is by
$A_{\mu}(s),  \partial_{\nu} A_{\mu}(s)$,
and agrees with $A_{\rho}$ on
the original links of the $b$-lattice. For low momentum this corresponds
to eq.(\ref{aq}).
 Overall momentum conservation is violated by multiples of
$2 \pi/b$ as is familiar in lattice theories in general, the difference in the
present context being that the fermion momenta live in the $f$-lattice
Brillouin zone not the $b$-lattice Brillouin zone. This is not a problem
for locality but is related to taking the $f \rightarrow 0$ limit,
 which is dealt with  in the next section.

$\bullet$ Rotational and translational symmetry:
Our interpolation scheme is manifestly invariant under rotations and
translations of the $b$-lattice so any invariant functional of $a_{\mu}$
translates into an invariant functional of $U_{\mu}$.

$\bullet$ Topological Winding Number: As noted by 't Hooft \cite{thooft},
 this type of interpolation allows us to ascribe a
winding  number to any $b$-lattice configuration whose field strength
dies off at infinity, namely the winding number of its continuum
interpolation. This will be useful in section 7, dealing with fermion
number violation. Clearly, `instantons' smaller than $b$ cannot occur  as the
interpolation of a $b$-lattice gauge field.
The basic idea of assigning winding number to lattice gauge fields was first
explored in ref.
 \cite{luscher1}.

It is good  to see the simplest concrete example of  many of the above
considerations. This is
provided by the interpolation of $U(1)$ gauge fields, for which we can
explicitely construct the solution,
\begin{equation}
a_{1}(x = s + y b) = \sum_{z_{2}, z_{3}, z_{4} =0,1}
 A_{1}(s + b \sum_{\rho \neq 1} z_{\rho} \hat{\rho})
\prod_{\nu \neq 1} (z_{\nu}(2 y_{\nu} - 1) + 1 - y_{\nu}),
\end{equation}
the other components obtained by cyclic permutation.
It is easily checked that this interpolation satisfies all the right
boundary conditions, is in Lorentz gauge and satisfies Maxwell's equations
(minimizes the Maxwell action for the appropriate dimension) in the interiors
of hypercubes, 3-cubes and plaquettes, and is transversely but not
longitudinally continuous.
Also, $b$-lattice gauge fields which are
gauge equivalent interpolate to gauge equivalent fields in the continuum or
$f$-lattice.

It is also instructive to examine the interpolation in momentum space,
\begin{equation}
a_{\mu}(q) = A_{\mu}(\overline{q}) \frac{e^{i b \overline{q}_{\mu}} - 1}{i (b
\overline{q}_{\mu} +
2 \pi m_{\mu})} \prod_{\nu \neq \mu}
\frac{2 - 2 \cos b \overline{q}_{\nu}}{(b \overline{q}_{\nu} +
2 \pi m_{\nu})^2},
\end{equation}
where we decompose continuum ($f$-lattice) momenta according to
\begin{equation}
q_{\mu} = \overline{q}_{\mu} + \frac{2 \pi}{b} m_{\mu} {\rm ~~~where~~}
 |\overline{q}_{\mu}| < \pi/b, ~~~{\rm and}~~m_{\mu}~
{\rm is~ an ~integer}.
\end{equation}
We see that $a_{\mu}$ has momentum components above $\pi/b$ ($m_{\mu} \neq
0$), though $A_{\mu}$ does not. But because of the transverse continuity of
the interpolation, above $\pi/b$ the momentum components fall off (with
$m_{\mu}$), faster
in the transverse directions than in the longitudinal direction. This
type of falloff will be sufficient for our purposes in the next section. At low
momentum, $q_{\nu} b \ll 1$, the interpolation looks very simple, $a_{\mu}(q)
\sim A_{\mu}(q)$. Finally, note that the interpolation can be expanded as a
power series in $b \overline{q}_{\mu}$
 so that there is no non-locality introduced
by substituting the interpolation into a local interaction vertex of
$a_{\mu}(q)$ to fermion fields.

\section{The $f\rightarrow 0$ Limit}
We have seen in section 2 that $\Gamma_r[u]$ at one loop
is finite in the limit $f\rightarrow 0$, and gauge invariant up to $O(f^2)$.
 Also, it was shown that $G[u]$ is gauge covariant up to $O(f^2)$.
However, when $\Gamma_r$ and $G$ are considered as functionals of the real
dynamical variables, $U$, since
the interpolation $u[U]$ depends on $f$, we must take care that
no extra powers of $1/f$ are generated.
The objective of this section is then to prove,
\begin{eqnarray}
lim_{f\rightarrow 0} \;\;\;  \Gamma_r[u[U]] < \infty, \;\;\;
lim_{f\rightarrow 0} \;\;\;  G[u[U]] < \infty,
\label{finiteness}
\end{eqnarray}
and
\begin{eqnarray}
\Gamma_r[u[U^\Omega]] - \Gamma_r[u[U]] = O(f^2),  \nonumber\\
G^{\omega^{-1}}[u[U^\Omega]] - G[u[U]] = O(f^2),
\label{gaugeinv}
\end{eqnarray}
where $\Omega(s)$ is a gauge transformation on the $b$-lattice, while
$\omega[U,\Omega]$ is its interpolation
to the $f$-lattice. (Recall that our interpolation has the property
that $u[U^\Omega] = u^\omega[U]$.)~
 $G^{\omega^{-1}}$ is the obvious gauge transformed
fermion propagator.

\subsection{Fermionic Effective Action}

In perturbation theory, the renormalized one loop effective action in terms
of the interpolated fields is given by,
\begin{eqnarray}
\Gamma_r[u] = \sum_n \int_{BZ(f)} \prod_i d^4 q_i
\Gamma_{\alpha_1 \alpha_2 ...}^{(n)}(q_1,..,q_n;f)
\;\; a_{\alpha_1}(q_1) ...a_{\alpha_n}(q_n) \; \delta^{(4)}(\sum_i q_i)
\end{eqnarray}
which can be rewritten as,
\begin{eqnarray}
\Gamma_r = \sum_n \int_{BZ(b)} \prod_i d^4 \bar{q}_i
\sum^{b/f}_{m_{\mu_i}=-b/f} \Gamma_{\alpha_1 \alpha_2 ...}^{(n)}(q_1,..,q_n;f)
a_{\alpha_1}(q_1) ...a_{\alpha_n}(q_n) \delta^{(4)}(\sum_i q_i)
\label{gammaeff}
\end{eqnarray}
where $q_i = \bar{q}_i + 2\pi m_i/b$ and $m_{\mu_i}= -b/f,..,b/f$.

Although the amplitudes $\Gamma^{(n)}$ are finite
in the limit $f \rightarrow 0$,
it is clear from eq. (\ref{gammaeff}) that the sums over integers
$m_{\mu_i}$
could introduce powers of $1/f$.
Obviously, for a smooth enough interpolation, the high momentum
components of the interpolated fields are expected to be cut off by $\sim 1/b$,
in such a way that
the sums over $m_{\mu_i}$ are convergent in the limit $f\rightarrow 0$.
We now prove that this is indeed the case. For the proof we only need some
smoothness properties of the interpolation and a bound on
one loop (fermion) lattice integrals for large external momenta $q\sim 1/f$.

In the previous section we constructed a tranversely continuous interpolation.
The gauge field
however jumps in the longitudinal direction at the boundaries of the
$b$-cell, i.e. $a_\mu(n)$ is not continuous in the $\mu$ direction, but it is
 in any tranverse direction
$\nu \neq \mu$.  Also, within each $b$-cell it is fully continuous with
continuous derivatives to arbitrary order. By continuity on the lattice
we mean the \
following property,
\begin{eqnarray}
a_\mu(n) - a_\mu(n-1) = O(f).
\label{cont}
\end{eqnarray}
Note that this definition already implies the existence of a derivative that
maybe discontinuous, but is bounded for small $f$.
For an interpolated field with these properties, it
 is easy to derive a bound on the high momentum components,
\begin{eqnarray}
|a_\mu(q)| \;\; < C \frac{1}{|\hat{q}_\mu|} \prod_{\alpha\neq \mu}
\frac{1}{|\hat{q}_\alpha^2|}  < C' \frac{1}{|q_\mu|} \prod_{\alpha\neq \mu}
\frac{1}{q_\alpha^2},
\label{formfactor}
\end{eqnarray}
with
\begin{eqnarray}
\hat{q}_\mu \equiv \frac{e^{iq_\mu f} -1}{f}.
\end{eqnarray}
where $C, C'$ are constants independent of $f$. Note that this bound is
satisfied by the $U(1)$ interpolation of the previous section.

We sketch the proof for a function of one variable, the extension being
straightforward. A function, $F(n)$, defined on the points $n f$, which
is not continuous at the boundaries $n f = s b$ (but continuous
with continuous derivatives elsewhere) can always be
written as the sum of a step-like function $S$ (i.e. constant in each
interval $(s-1) b \leq n f < s b$) plus a continuous function $\tilde F$,
\begin{eqnarray}
F(n) = \tilde{F}(n) + S(s \frac{b}{f}),
\end{eqnarray}
with $n f = s b + \delta n f$, and $0 \leq \delta n < b/f$.
The momentum components of $S$ behave as $O(1/|{\hat q}|) < C/|q|$, for
$q\sim 1/f$. On the other hand, since $\tilde{F}$ is continuous in the
sense of (\ref{cont}), its first lattice derivative is bounded in the
limit $f\rightarrow 0$. Also the momentum components of this derivative
are bounded and we can write,
\begin{eqnarray}
|{\tilde F}(q)| = |\frac{({\hat \partial}{\tilde F})(q)}{\hat{q}}| <
 C \frac{1}{|q|}
\end{eqnarray}
This implies  that $|F(q)| < C/|q|$, with
$C$ independent of $f$.
 It is easy to repeat the analysis for a continuous function, by
writing its derivative as the sum of a step-like function and a continuous
function. We find in this case $|F(q)| < C/|q|^2$.

The second ingredient in the proof of finiteness is a simplified
version of the power counting theorem of Reisz \cite{reisz}. We
use the derivation of reference \cite{luscher2} for one loop diagrams.
Here, we simply state the result and refer to appendix A for the details.
The renormalized lattice
integrals contributing to the $\Gamma^{(n)}$ of (\ref{gammaeff})
 satisfy the bound
\begin{eqnarray}
|\Gamma^{(n)}(q;f)| \;\; < \;\; C |q|^{4-n} Logs(|q|),
\label{lusher}
\end{eqnarray}
for all $f < f_0$ for some fixed $f_0$, with $C$ a constant
independent of $f$. $n$ is the number of external boson lines
and $q$ denotes generically the external momenta. Note that this is also
the result expected from naive power counting in the continuum.

Now, we are ready to prove the finiteness of (\ref{gammaeff}). First
we note that for $n \geq 4$, eqs. (\ref{formfactor}) and (\ref{lusher})
imply that the sums over $m_\mu$  are absolutely convergent in the limit
$f\rightarrow 0$. Since
the genuine logarithmic divergence in the $n =4$ amplitude has been absorbed
into the definition of $g_b$ (see eq.(\ref{coupling})), this means that all
the contributions with $n\geq 4$ are finite in this limit.

To prove finiteness in the $n=2$ and $n=3$ cases, we have to rely on
the transverse properties of these amplitudes for the hybrid action.
The P-even contributions are gauge invariant. This implies,
\begin{eqnarray}
\sum_\mu \hat{q}_\mu \Gamma^{(2)}_{\mu\nu}(q) = 0  \;\;\;\;
\sum_\mu \hat{q}_{1,2}^\mu \Gamma^{(3)}_{\mu\nu\alpha}(q_1,q_2) =
C\; \Gamma^{(2)}_{\nu \alpha}(q_{2,1}).
\label{gipe}
\end{eqnarray}
It is shown in Appendix A that these relations are enough to prove the stricter
bounds,
\begin{eqnarray}
|\Gamma^{(2)}_{\mu\mu}(q)| < C |q_\alpha|^2 Log(|q|) \;\;\;\;
|\Gamma^{(3)}_{\mu\mu\mu}(q)| < C |q_\alpha| Log(|q|) \;\;\;\;\alpha \neq
\mu.
\label{trans}
\end{eqnarray}
Together with (\ref{formfactor}) and (\ref{lusher}), it is now
straightforward to prove absolute convergence of the $m_\mu$ sums for
 the parity-even $n=2$ and $n=3$.

On the other hand, these relations are not satisfied by the
$\chi w$ action (\ref{cw}) and
in fact we cannot prove convergence of the $m_\mu$ sums in
 (\ref{gammaeff}). This is simple to understand
in real space. We have seen that the chiral Wilson action requires a
counterterm of the type,
\begin{eqnarray}
\delta \chi(f) \;Tr\;\sum_{\mu} ( \Delta_\mu u_\mu )^2.
\end{eqnarray}
This is not well defined in the $f\rightarrow 0$ limit, because the
longitudinal derivative of the interpolated
field is a sum of approximate delta functions, whose square is ill-defined.
For finite $f$, it would diverge as a power of $b/f$, introducing more
divergences than those in $Z_W$.
In the absence of a smoother interpolation, we conclude that the chiral
Wilson action (\ref{cw}) cannot be used.

Finally, we need to consider the parity-odd parts in $n=2,3$. The P-odd
amplitude for $n=2$ is exactly zero. On the other hand, the three-point
function is not zero.
Obviously, since anomalies cancel, these contributions are gauge invariant in
the limit $f\rightarrow 0$, but for finite $f$, we cannot use the fact
that the diagram is gauge
invariant to derive a relation like (\ref{trans}). Nevertheless,
this bound still holds because of the presence of an epsilon
tensor. It is proven in appendix A by direct
inspection. Again this is enough to prove absolute convergence
of the $m_\mu$ sums.

This concludes the proof that the one-loop renormalized fermionic effective
action derived for the hybrid scheme (\ref{hs}) has a good limit $f
\rightarrow 0$.

We still need to prove that the gauge non-invariant contributions that
remain after renormalization are $O(f^2)$. In section 2, we have seen that
this is true for $\Gamma_r[u]$.
Here we prove that, when the $f$ dependence
of the interpolation is taken into account in
$\Gamma_r[u[U]]$, no extra powers of $1/f$ change this result
(\ref{gaugeinv}). For this
we just need to extend the bounds (\ref{lusher}) to amplitudes with one
insertion of $\delta_\omega S^{\chi w}_f$. We denote these amplitudes
by $\delta_\omega \Gamma^{(n)}$. Under a gauge transformation on the
$b$-lattice, $\Gamma_r$ for the hybrid scheme changes by,
\begin{eqnarray}
\delta_\omega \Gamma_r = \delta_\omega
\ln(\frac{\det i\hat{D}}{\det [i\hat{D}]^*})^{\frac{1}{2}}
= Im[< \delta_\omega S^{\chi w}_f >_{\chi w}] =\nonumber\\
\sum_n \int_{BZ(b)} \prod_i d^4\bar{q}_i \sum^{b/f}_{m_{\mu_i}=-b/f}
\delta_\omega \Gamma^{(n)}(q_i) a(q_1) a(q_2)...a(q_n) \delta^{(4)}(\sum_i
q_i),
\label{deltaG}
\end{eqnarray}
where, $S_f^{\chi w}$ is given by  the action (\ref{cw}).
In momentum space, we get
\begin{eqnarray}
\delta_\omega S^{\chi w}_f = i \theta^a(q) \delta^{(4)}(q+p-p')
\bar{\Psi}(p') \{ M(p') L - M(p) R \} T^a \Psi(p),
\label{ins}
\end{eqnarray}
for an infinitesimal gauge transformation $\omega=e^{i \theta^a(n) T_a}$,
obtained by interpolating an infinitesimal $b$-lattice gauge transformation
$\Omega(s)$ (see section 3), and $M(p) \equiv (2 r/f) \sum_\nu \sin^2(p_\nu
f/2)$. Inserting this
term into a free fermion propagator gives,
\begin{eqnarray}
- i \;[ \frac{M(p')^2}{P_f^2(p')} \frac{1}{P_f(p)} - \frac{M(p)^2}{P^2_f(p)}
\frac{1}{P_f(p')} ],
\end{eqnarray}
where $1/P_f$ is the free fermion propagator. Also, we have the bound,
\begin{eqnarray}
|\frac{1}{f^2} \frac{M^2(p)}{P^2_f(p)}| < C p^2,
\label{inbound}
\end{eqnarray}
for all $f < f_0$ and $C$ some constant. From the analysis of
Appendix A, we conclude that a diagram
with one insertion of $\delta_\omega S_f$ and
$n$ gauge legs, is the sum of local terms up to dimension $5$ plus a
non-local part which is bounded by,
\begin{eqnarray}
|\delta_\omega \Gamma^{(n)}| \leq f^2 |q|^{6 -n} Logs(|q|).
\label{boundf2}
\end{eqnarray}
Since there is no local P-odd operator of dimension 5 or less, except
the anomaly, which vanishes, the non-local part bounded by (\ref{boundf2}) is
 all there is
and as expected is $O(f^2)$.

It is clear from the previous discussion that amplitudes with $n \geq 6$
have two or more powers of $f^2$ and since for large momentum, $q\sim 1/f$,
they must satisfy bound (\ref{boundf2}), we can use it together with
(\ref{formfactor}), to prove that the sums over $m_\mu$ in (\ref{deltaG}) are
absolutely
convergent and do not bring any extra powers of $1/f$.

For $n \leq 5$, we make use of the fact
that there is an epsilon tensor. In the case of $n=2$, it can be shown by
direct inspection that the integral is bounded by,
\begin{eqnarray}
|\delta_\omega \Gamma_{\mu\nu}^{(2)}(q)| < C
f^2 |q_\alpha| |q_\beta| |q|^2 Log(|q|) \;\;
\alpha,\beta \neq \mu,\nu \;\;\; \alpha\neq \beta
\end{eqnarray}
Using (\ref{formfactor}) and the full continuity of the interpolated gauge
transformation $\theta(n)$,
which leads to a bound of the form
\begin{eqnarray}
|\theta(q)| \;\; < C \prod_{\alpha} \frac{1}{q_\alpha^2},
\end{eqnarray}
it is easy to see that
the sums over $m_\mu$ yield a result suppressed by $f^2$ up to logs.
A similar analysis shows this is true for  $n=3-5$ diagrams.
This ends the proof of (\ref{gaugeinv}).

\subsection{Fermion Propagator}
We finally need to consider the propagator $G[u[U]]$. We will prove that it has
a well-defined $f \rightarrow 0$ limit and
that the breakage of gauge invariance is down by at least $O(f^2)$.
In perturbation theory, it is given by,
\begin{eqnarray}
G[U](q,q') = \sum_n \int_{BZ(b)} \prod^{n}_{i=1} d\bar{p}_i \sum_{m_{\mu_i}}
G^{(n)}(p_1,..,p_n)
a(p_1)...a(p_n) \delta^{(4)}(q'-q-\sum^{n}_{i=1} p_i)
\label{propa}
\end{eqnarray}
where as usual $p_i = \bar{p}_i + 2 \pi m_i/b$.
Since, the amplitudes $G^{(n)}$ are products of free propagators, and
$|P_f(k)| < C/|k|$, this implies that for large $m_\mu$,
we must have $|G^{(n)}| < C/|p_i|$, for any large $p_i$. This bound
together with (\ref{formfactor}) implies the
absolute convergence of the sums over $m_{\mu}$ in (\ref{propa}).
The $f\rightarrow 0$ limit of $G[U]$ is well-defined.

Now we have to prove that the gauge non-invariant contributions to these
diagrams are $O(f^2)$. The gauge variation of $G[U]$ is given perturbatively
by a sum of diagrams with an arbitray number of gauge boson legs and one
insertion of $\delta_\omega S_f^{\chi w}$. Using (\ref{ins}) and
(\ref{inbound}), we see that the line with the insertion
can be bounded for large $m_\mu$ by
$f^2 |k|$ where $k$ is the momentum of the fermion line,
 as compared to the behaviour $\sim 1/|k|$ without the insertion. From
(\ref{propa}), it is clear that the sums over $m_\mu$ are at most
logarithmically divergent in this case, so that the
breakeage of gauge invariance in $G[U]$ is at most $O(f^2 Log(f))$,
as expected.

\section{Gauge Ward Identities.}

We are now in a position to demonstrate the gauge invariance of our
construction
when gauge fields, $U$, are integrated in the functional integral.
We have shown in the previous section that
the three components of the full functional integral (see eq.
(\ref{effact})), $\Gamma_r[u[U]], G[u[U]], S_g[U]$, satisfy gauge invariance
up to $O(f^2)$ (in fact we showed that mathematically the gauge invariant limit
 $f \rightarrow 0$ exists, but in any simulation $f$ will be small and
non-zero). In the absence of gauge fixing, the full generating functional
satisfies,
\begin{eqnarray}
&~&\int \prod_s d U_{\mu}(s)~ e^{- S_{g}[U] + \sum_s J_{\mu}
A^{\Omega}_{\mu}} ~e^{\Gamma_r[u[U^{\Omega}]]} ~ e^{-\overline{\eta}
G^{\omega^{-1}}[u[U^{\Omega}]] \eta} = \nonumber
\\
&~&\int \prod_s d U_{\mu}(s)~ e^{- S_{g}[U] + \sum_s J_{\mu}
A_{\mu}} ~e^{\Gamma_r[u[U]]} ~ e^{-\overline{\eta}
 G[u[U]] \eta} + O(\frac{f^2}{b^2}),
\label{ward}
\end{eqnarray}
where $\Omega$ is any
$b$-lattice gauge transformation while $\omega[\Omega,U]$ is its $f$-lattice
interpolation described in section 3. The gauge transformed propagator is given
by,
\begin{eqnarray}
G^{\omega^{-1}}[u[U^{\Omega}]](n_1, n_2) \equiv \omega^{-1}(n_1)
G[u[U^{\Omega}]](n_1, n_2)
 \omega(n_2).
\end{eqnarray}
The central observation needed to understand eq.(\ref{ward}) is
that since the $U$ fields
live on the $b$-lattice, the largest momentum scale they can
introduce when integrated is $O(1/b)$. That is, they cannot reintroduce any
powers of $1/f$ which could cancel the $f^2$ suppression of gauge symmetry
violation. Note that this is a non-perturbative expectation
 \footnote{Small deviations from exact gauge invariance, such as a
small bare gauge boson mass term, lead to difficulties in continuum
perturbation theory, as is well known. Indeed one is really outside the
perturbative regime. However non-perturbatively there is no sign of
difficulty. For a brief review of the lattice literature on the example of bare
gauge boson masses see ref. \cite{smit3}.}.

For perturbative investigations it is useful to add the standard,
rotationally invariant gauge-fixing and ghost terms, neglecting the small
breakage of gauge invariance in $\Gamma_r[u[U]]$ and $G[u[U]]$ in the
Fadeev-Popov procedure. This results in lattice BRST identities among the
Green functions which are broken only by $O(f^2/b^2)$, since the gauge
boson loop momenta are cut off above $O(1/b)$.
The statement of the gauge invariance and unitarity of the
quantum theory is contained in the  BRST identities. From these
it should also be possible to prove perturbative
renormalizability and the continuum (renormalized) BRST identities
of the theory. Of course, it is believed that
there is a unique
 continuum theory with the particle content of our lattice construction
 which satisfies the BRST identities pertubatively, and non-perturbatively
coincides with the non-gauge-fixed theory.

Note that to compute physical amplitudes, it is essential to be
able to form gauge invariant combinations of the external fermion fields
and the gauge fields. This, at first sight,  seems problematic  since the
fields live
on different lattices, but clearly the solution is to make use of the
gauge field interpolation $u[U]$ in forming path-ordered exponentials
of the gauge field connecting external fermions.

Let us turn to the question of how small to take the ratio $f/b$ in a
lattice simulation in a (physical) volume, $L^4$. Clearly we should take
$f/b$ small enough that the measurements in our volume are insensitive to
the breakage of gauge invariance. For example, a
 simple perturbative estimate suggests that (in the gauge-fixed theory) the
$O(f^2/b^2)$ breaking of the Ward identities can induce a gauge boson
mass of order $f/b^2$. Clearly, if
$f/b < b/L$  we would be insensitive to this violation of gauge invariance
in our volume, since the gauge boson correlation length would be larger
than $L$. With such a relationship of scales, in
 the continuum or thermodynamic limits, $b/L \rightarrow 0$,
gauge invariance becomes exact ($f/b \rightarrow 0$).

\section{Anomalies in Gauge and Global Currents}
In this section we will first check that our hybrid scheme reproduces
the consistent gauge anomaly at the one-loop level.
As explained in section 2, this is to be expected on general grounds.
Obviously, we only consider theories where the gauge anomaly vanishes (i.e.
$d_{abc}$ among the fermion
representations is zero). However, in most theories of interest, there
are also other currents associated with global symmetries that can be
anomalous (e.g. B+L number in the Standard Model). We will show that these are
also properly reproduced. In studying current conservation, we
will need to consider Green functions with one insertion of the global
current. In general, new counterterms are needed to
ensure gauge invariance of amplitudes containing composite operators such
as this current, even in the hybrid scheme.
Once these counterterms have been added,
we show that our construction gives the correct gauge invariant
anomaly at the one-loop level.
It is then straightforward to see that the Adler-Bardeen  theorem
\cite{adler}
must hold to all orders in perturbation theory.

 The axial anomaly triangle in a vector-like theory was computed in lattice
perturbation theory
with standard Wilson fermions by several authors \cite{smit}\cite{kawai}.
The case we have to consider here is slightly different, since we are in
a chiral gauge theory \footnote{For a discussion of gauge and global anomalies
in the context of the
`overlap' formulation of chiral
gauge theories see ref. \cite{overlap}}. The gauge anomaly triangle for our
hybrid scheme is
given by the imaginary part of the diagram with one insertion of
(\ref{ins}) and two gauge boson legs, in the limit $f\rightarrow 0$ (recall
that in the hybrid scheme only $Im\; \Gamma$ breaks the gauge symmetry).
The result we get is,
\begin{eqnarray}
\delta_\omega {\Gamma_{\mu\nu}^{(2)}}^{b c}(q,q')  =
\frac{i}{\pi^4} \theta_{a}(q+q') d_{abc} \epsilon_{\mu\alpha\nu\beta}
q'_\alpha q_\beta \times \nonumber\\
\int^{\pi}_{-\pi} d^4k\; c_\mu c_\alpha c_\nu s_\beta^2\; \frac{r M s^2 -
c_\beta M^2 }{(s^2+M^2)^4} + O(f^2)
\end{eqnarray}
where $\omega = exp(i \theta^a T^a)$. Also $c_\alpha \equiv cos(k_\alpha)$,
$s_\alpha \equiv sin(k_\alpha)$,
$M \equiv 2 r \sum_\alpha sin(k_\alpha/2)^2$ and $s^2 \equiv \sum_\alpha
sin(k_\alpha)^2$. $r$ is the Wilson parameter.
This  integral can be solved exactly \footnote{We thank Ph.Boucaud for
pointing this out to us.} and gives a result independent of
$r$ as expected,
\begin{eqnarray}
\delta_\omega \Gamma^{(2)}(q,q')  =
 \; \frac{i}{12\pi^2} \theta^{a}(q+q') d^{abc} \epsilon_{\mu\alpha\nu\beta}
 q'_\alpha q_\beta
+ O(f^2).
\end{eqnarray}
The Wess-Zumino consistency conditions \cite{zumino}
then give the second term of the gauge anomaly,
\begin{eqnarray}
\delta_\omega \Gamma  = - \frac{i}{12 \pi^2} \epsilon_{\alpha\mu\beta\nu}
Tr[ \theta\; \partial_\alpha ( a_\mu \partial_\beta a_\nu - \frac{i}{2}
a_\mu a_\beta a_\nu ) ] + O(f^2).
\label{gaugeano}
\end{eqnarray}
We have proven earlier that, in the absence of gauge anomalies,
the hybrid scheme gives a gauge symmetric continuum limit.
However, the chiral gauge theories we want to simulate typically contain
global $U(1)$'s which are anomalous.
We will consider a general case, in which a global current is denoted
by $j_\mu$, with generator $t$. We will assume that $t$
commutes with the $T_a$.
A convenient way to find the anomaly is introducing an external source
$b_\mu(x)$ that couples to the global current. It also lives on the links of
the $f$-lattice: ${\tilde u}_\mu = exp(i f b_\mu t)$.
The simplest choice is  to couple it like the gauge field (any other
choice would lead to the same result). In the hybrid scheme we
just substitute,
\begin{eqnarray}
u_\mu \rightarrow u_\mu {\tilde u}_\mu
\end{eqnarray}
both in the chiral Wilson action (\ref{cw}) and in the standard Wilson
action, defining $\det[i\hat D]_{\chi w}$ and $\det[i\not\!\!D]_{sw}$
in (\ref{hs}).

Note that for the chiral Wilson action this prescription corresponds
to only the
physical left-handed fermion field transforming under the global symmetry.

Due to the presence of the external field $\tilde u$, the fermionic effective
action is no longer gauge invariant for $d_{abc} =0$. Instead we get,
\begin{eqnarray}
\delta_\omega \Gamma[u,\tilde u] = - \frac{i}{12 \pi^2}
\epsilon_{\alpha\mu\beta\nu} Tr[\theta \partial_\alpha ( 2\;a_\mu
\partial_\beta b_\nu - \frac{i}{2} b_\mu a_\beta a_\nu )] + O(b^2_\mu) + O(f^2)
\end{eqnarray}
Since we are only interested in one insertion of the current, we just need
the linear terms in $b_\mu$. It is easy to see that by adding the local
 counterterm,
\begin{eqnarray}
S_{c.t.}[a,b] = \frac{i}{12 \pi^2}  \epsilon_{\alpha\mu\beta\nu}
Tr[2\; b_\alpha a_\mu \partial_\beta a_\nu - i \frac{3}{2} b_\alpha
a_\mu a_\beta a_\nu]
\end{eqnarray}
one recovers gauge invariance,
\begin{eqnarray}
\delta_\omega ( \Gamma + S_{c.t.} ) = O(f^2).
\end{eqnarray}

Finally, after this subtraction, the global anomaly is given by,
\begin{eqnarray}
\delta_\lambda (\Gamma + S_{c.t.}) \; |_{b_\mu=0} = Im( <\delta_\lambda
S_F^{\chi w} >_{\chi w} )|_{b_\mu =0}\; + \delta_\lambda S_{c.t.}\; |_{b_\mu
=0}
\end{eqnarray}
where $\lambda = e^{i \phi(n) t}$.
The first term corresponds to a diagram which is identical to the one for
the gauge anomaly,
with $\phi\; t$ replacing $\theta_a T_a$ in (\ref{gaugeano}),
\begin{eqnarray}
\delta_\lambda \Gamma\;|_{b=0} = - \frac{i}{12 \pi^2} \phi\;
\epsilon_{\alpha\mu\beta\nu} Tr[t\; \partial_\alpha (
 a_\mu \partial_\beta a_\nu - \frac{i}{2} a_\mu a_\beta a_\nu ) ] + O(f^2)
\end{eqnarray}
When we add the counterterm we find,
\begin{eqnarray}
\phi\;\partial_\mu j_\mu \equiv
\delta_\lambda ( \Gamma + S_{ct} )\; |_{b_\mu=0} =
- \frac{i}{16 \pi^2} \phi\; Tr[t\; f_{\mu\nu} {\tilde f}_{\mu\nu}] + O(f^2)
\label{anof}
\end{eqnarray}
with $f_{\mu\nu}$ being the gauge field strength. This is the expected gauge
 invariant
form of the anomaly.

This is the result at one loop. However, since gauge loops can at most
 generate extra powers of $1/b$,
all the $O(f^2)$ effects in (\ref{anof}) go to zero in the continuum limit,
$f/b \rightarrow 0$, to all orders in
perturbation theory. This means that in the continuum limit, we recover the
well-known result of the Adler-Bardeen theorem since,
\begin{eqnarray}
< \partial_\mu j_\mu \;F > ~=~
 < -\frac{i}{16 \pi^2} Tr[t f_{\mu\nu} {\tilde f}_{\mu\nu}] \;F > +
< \delta_t F >  + O(f^2/b^2)
\end{eqnarray}
where $F[U,\Psi]$ is an arbitrary functional.

\section{Fermion Number Violating Amplitudes.}

Consider the example of a theory with one generation of standard model
fermions interacting with gauge fields but no scalars (since we have not
considered them in this paper). In this section we will understand
the phenomenon of $B+L$ violation \cite{thooft2} in an `instanton gas'
picture, in the
continuum and then on the lattice.
This will be sufficient to
demonstrate that $B+L$-violating processes will occur in our lattice
construction. We then give a prescription for defining lattice-regulated
net-$B+L$ violating amplitudes. A similar discussion has been previously
given in reference
\cite{overlap}, from the point of view of the `overlap' formulation of
chiral gauge theories.

The main problem is that we have restricted the  lattice
construction to net fermion number conserving amplitudes (for every external
$\Psi_L$ we have a corresponding
external $\overline{\Psi}_L$). Still, as we saw in the last section, even
in this sector of the theory there is {\it local} $B+L$ violation,
as given by the anomaly. This will allow us to infer
$B+L$ violation in the full theory.
The strategy for defining more general
correlators, say $<{\cal O}(x)>$ representing the amplitude for a
$B+L$-violating process in the vicinity of $x$, is to demand cluster
decompostion of the full theory and consider instead the
fermion number {\it conserving} correlator \cite{rome2},
\begin{equation}
 <{\cal O}(x) {\cal O}^\dagger(y)>_{|x - y| \rightarrow \infty}\;\; ~\sim ~
<{\cal O}(x)> <{\cal O}^\dagger(y)>,
\end{equation}
which is defined in our construction.
That is, if this two-point correlator does not vanish as $x$ and $y$ are
taken arbitrarily far apart, then $B+L$ is being {\it locally} violated (with
opposite signs) at $x$ and $y$. Since events at $y$ are too distant to
affect those at $x$ in a sensible theory, we must have that
\begin{equation}
<{\cal O}(x)> \neq 0.
\end{equation}
Inferring fermion number violation by considering fermion number conserving
correlators has been exploited in refs. \cite{ringwald}.

We will first see how this works formally in the continuum, using the
formalism of Fujikawa for chiral theories \cite{fuji},
 and then consider our lattice construction afterwards.
For simplicity, we will take
\begin{equation}
{\cal O}(x) = q_L(x) q_L(x) q_L(x) \ell_L(x),
\end{equation}
where we take the internal indices to be contracted gauge invariantly. This
operator
carries two units of $B+L$. We then have
\begin{equation}
{\cal O}^{\dag}(y) = \overline{q}_L(y) \overline{q}_L(y) \overline{q}_L(y)
\overline{\ell}_L(y).
\end{equation}
For a fixed background gauge field $a_{\mu}$, we
can decompose the chiral quantum fields  in terms of the
complete, orthonormal set of eigenfunctions of the {\it vector-like} hermitian
Dirac operator, $i\not\!\!D$,
\begin{eqnarray}
 i\not\!\!D \psi_n &=& \lambda_n \psi_n \nonumber \\
\Psi_L &=& \sum_{n > 0} a_n \psi_n^L \nonumber \\
\overline{\Psi}_L &=& \sum_{n > 0} \overline{a}_n \psi_n^{R \dagger}.
\end{eqnarray}
 Here, we are using a condensed notation where really there are separate
Dirac equations for the $\Psi_L = q_L, \ell_L$ fields. The two-point
correlator of ${\cal O}$ is then given by functional integration over the
fermion coefficients $\overline{a}_n, a_n$,
\begin{equation}
<{\cal O}(x) {\cal O}^\dagger(y)>_{a_{\mu}} = (\prod_{n > 0} \lambda^q_n)^3
(\prod_{n > 0} \lambda^{\ell}_n)
e^{i {\rm Im} \Gamma} (\sum_{n>0} \frac{q_n^L(x) \overline{q}_n^{R
\dag}(y)}{\lambda^q_n})^3
(\sum_{n>0} \frac{\ell_n^L(x) \overline{\ell}_n^{R
\dag}(y)}{\lambda^{\ell}_n}),
\end{equation}
where $<...>_{a_{\mu}}$ denotes the fermionic averaging done in the
background of $a_{\mu}$, which is  presently unintegrated.
The origin of the various terms is as follows. The products of eigenvalues
for the various fields comes from integrating the fermion action, which is
diagonalized in this basis. This is nothing but the square root of the
vector-like determinant, which we gauge-invariantly regulated in section 2.
 The phase factor emerges as a Jacobian for a formally unitary
transformation,
\begin{equation}
\int {\cal D} \Psi_L {\cal D} \overline{\Psi}_L = \int \prod_{n>0} da_n
d \overline{a}_n e^{i {\rm Im} \Gamma}.
\end{equation}
The bilinears in the eigenfunctions are the fermion propagators in the
eigenbasis.

Now we specialize to the case of a zero winding number background gauge
field which is made out of a distantly separated instanton and
anti-instanton pair, $a_{\mu} + \tilde{a}_{\mu}$, the instanton in the
vicinity of $x$ and the anti-instanton in the vicinity of $y$.
To understand the dominant contribution to our two-point correlator we
note that for a single instanton, the Atiyah-Singer
index theorem \cite{atiyah}
tells us that each of the fields comprising ${\cal O}$ has a left-handed
zero mode of its associated vector-like Dirac eigen-equation. For an
anti-instanton background, there are right-handed zero modes of the Dirac
eigen-equation instead. In the background of the instanton anti-instanton
pair, the left and right-handed zero modes pair up to form a Dirac
spinor eigenfunction with non-zero eigenvalue, $\lambda_1$. As the
instanton and anti-instanton are moved far apart, $\lambda_1 \rightarrow
0$ and the two chiralities separate spatially. As this happens
the fermion propagators are dominated by the $\lambda_1$ modes,
\begin{equation}
\sum_{n>0} \frac{\psi_n^L(x) \overline{\psi}_n^{R \dag}(y)}{\lambda_n}
{}~\sim~ \frac{\psi_1^L(x) \overline{\psi}_1^{R \dag}(y)}{\lambda_1},
\end{equation}
while the product of eigenvalues is suppressed by $\lambda_1$. Cancelling
the powers of $\lambda_1$ gives
\begin{equation}
<{\cal O}(x) {\cal O}^\dagger(y)>_{a_{\mu} + \tilde{a}_{\mu}}~
 \sim (\prod_{n > 1} \lambda^q_n)^3
(\prod_{n > 1} \lambda^{\ell}_n)
e^{i {\rm Im} \Gamma} (q_1^L(x) \overline{q}_1^{R
\dag}(y))^3
(\ell_1^L(x) \overline{\ell}_1^{R \dag}(y)),
\end{equation}
where the leading behaviors of $\psi_1^L$ and $\overline{\psi}_1^{R \dag}$
are given by the zero modes of the instanton and anti-instanton respectively.

The mechanism by which $<{\cal O} {\cal O}^{\dag}>$  survives
 as $|x-y| \rightarrow \infty$,
when we integrate over gauge fields in a dilute instanton gas
approximation, is now clear. There will be a purely perturbative
contribution to the correlator which will fall to zero as $|x-y| \rightarrow
\infty$ and a contribution from an instanton anti-instanton pair of the
above form which does not fall off, but rather factorizes into
 contributions from the vicinities of $x$ and $y$. We are
forced to conclude that $B+L$ violation occurs in the full theory. Of
course, we expect the qualitative point we are making to survive the
(rather drastic) dilute gas approximation.
The moral is that the physics of non-trivial
winding number is implicit in the zero winding number sector and this is
linked to $B+L$ violation through the (integral of the)
anomaly equation for the $B+L$ current.

Let us now see how all this relates to our lattice construction.
As described in section 3 we mean by a single
`instanton', a $b$-lattice configuration whose interpolation  to
the continuum yields a unit winding number continuum gauge field. This is
easily arranged by obtaining the original $b$-lattice gauge field by
discretizing a large (compared with $b$)  continuum instanton in the first
place. An instanton and anti-instanton are then easily composed to give a
zero winding number gauge field. For $f \ll b \ll$ instanton size,
 the hybrid regulator gives
essentially the same result as our
continuum analysis except that the product of eigenvalues is cut off above
roughly $1/f$. The vector-like determinant is determined by the gauge
invariant standard Wilson action. We can work with the related set of
lattice eigenfunctions, which for small $f$ we expect to approximate the
Fujikawa eigenfunctions of our continuum discussion.
The gauge field interpolation procedure makes the remaining analysis as
simple as the formal continuum case. The
fermion propagators are nearly the same as the continuum propagators in
the background of the interpolated gauge field. The poles corresponding to
the doubler modes have completely decoupled. Therefore the ${\cal O}$
two-point correlator is dominated by the approximate
 zero modes just as in the earlier
analysis. For $|x-y| \rightarrow \infty$ the correlator
 must be interpreted as $B+L$ violation occurring at $x$ and at $y$ but
without any physical connection between the two points.

The above discussion leads us to a straightforward means of defining
 arbitrary lattice-regulated fermion number violating amplitudes in terms of
the
fermion conserving amplitudes we have already defined. For example,
let $\tilde{\cal O}$ be a product of fields at various
points, which carries two units of $B+L$. Then for $x$ far removed from all
these points, cluster decomposition yields
\begin{equation}
<\tilde{\cal O} {\cal O}^{\dag}(x)>\; ~\sim~ \; <\tilde{\cal O}>
<{\cal O}^{\dag}(x)>.
\end{equation}
We can satisfy this by defining
\begin{equation}
<\tilde{\cal O}>~ \equiv  ~ \frac{<\tilde{\cal O}
{\cal O}^{\dag}(x)>}{(<{\cal O}(x){\cal O}^{\dag}(y)>)^{\frac{1}{2}}},
\end{equation}
for $y, x$ also distantly separated. Note that all the quantities on the
right-hand side
are defined in the zero-winding number sector\footnote{Actually, there is
 also an undetermined phase in $<{\cal O}>$, which we have taken as
zero. Its value in general defines the $\theta$-vacuum we occupy.}.

\section{Computational Efficiency.}

Obviously, the most computationally expensive operation in a simulation of
chiral gauge theory using the methods of this paper, compared with lattice
QCD say, is the calculation of the fermion determinant on the
$f$-lattice, instead of the coarser $b$-lattice. The other major complication
is
the gauge field interpolation technique.
 Let us see what can be done to reduce these costs.

In the hybrid scheme described in section 2, there are in fact two
determinants, the $sw$ and $\chi w$ ones, which must be calculated on the
$f$-lattice for each gauge field configuration. The consistent use of the
$f$-lattice for fermions was done just to reduce confusion.
It is however possible, and
much more efficient to
calculate the $sw$ determinant directly on the $b$-lattice! The point is that
 the $sw$ determinant is exactly gauge invariant and  we know how to
non-perturbatively
restore the global chiral symmetry broken by the standard Wilson term
 on a single lattice for gauge fields and fermions.
Just as in lattice QCD, we can use the action
\begin{eqnarray}
S_{sw} &=& \frac{b^3}{2} \sum_{s,\mu} \bar{\Phi}(s) \gamma_\mu U_\mu
\Phi(s+\hat{\mu})
 - \bar{\Phi}(s+\hat{\mu}) \gamma_\mu U^{-1} _\mu \Phi(s)\nonumber\\
&+& (4 r + M b) b^3\sum_s \bar{\Phi}(s) \Phi(s) \nonumber \\
&-&
\frac{r  b^3}{2} \sum_{s,\mu} [\bar{\Phi}(s) U_\mu(s) \Phi(s+\hat{\mu})
+\bar{\Phi}(s+\hat{\mu}) U_\mu^{-1} \Phi(s)],
\label{wpo}
\end{eqnarray}
where $M$ has been tuned to the point of chiral symmetry for the
 vector-like fermions $\Phi$, by tuning the associated Goldstone bosons to be
massless.  We see that in this formulation, there is
no need to do coupling constant renormalization for distances from $f$ to
$b$. That is,
\begin{equation}
g_b = g_0.
\end{equation}
The $\chi w$ determinant which is used to regulate Im$\Gamma$ must still be
regulated on the $f$-lattice because it is not exactly gauge invariant,
unlike the $sw$ determinant. Thus the workload has been cut in half, but not
more.

Similarly, one can use the $b$-lattice to regulate external fermions. If
a $\chi w$ action is used on the $b$-lattice, the breakage of gauge
invariance would not be suppressed. But instead, we can use the standard
Wilson action above, of course restricting ourselves to calculating with
the external fermions carrying the gauge charge.

For conceptual purposes, we have already considered matching the two-cutoff
theory with a one-cutoff theory. Here we note that such a matching would
result in our ability to compute in a chiral theory with the same
efficiency as in (unquenched) lattice QCD!
The Rome group proposal is similar to this one cutoff theory;
they employ the $\chi w$ regulator scheme
with gauge fixing terms and demand that BRST
relations  among the Green functions are obeyed in any
simulation, hoping to thereby tune the
 counterterm coefficients to all
orders. However, non-perturbatively the viability of this procedure is unclear.
On the other hand, one can imagine computing a handful of Green
functions using our two-cutoff method, possibly in a volume not too much
larger than $b^4$.
It may then be possible to  tune the counterterms in a single $b$-cutoff
theory in
order to reproduce the same results to good precision.
 Once the counterterms are determined in this
manner they are fixed for all other possible Green
functions. Therefore one can simply proceed to use the single cutoff
theory from then on. This corresponds precisely to a non-perturbative
matching having been performed at $b$ between the two-cutoff and one-cutoff
theories.

Next, we observe that the size of the breakage of gauge invariance is
partly determined by the form of the Wilson term in the $\chi w$ regulator. The
fact that it turned out to be $O(f^2/b^2)$ is directly linked to the fact
that the $f$-lattice Wilson term possessed two lattice derivatives. But nothing
prevents us from using a four-derivative Wilson term, for example. This
will push the unsubtractable non-local breakage of gauge invariance
to higher order in $f$. If all the lower dimensional local terms are still
locally integrable we could arrange to have the Ward identities obeyed to
higher precision than $f^2/b^2$, so a much coarser $f$-lattice could be used.
 There is a
problem however in that the interpolation will almost certainly need to be
smoother than it presently is.
 For example, a four-derivative counterterm would be
very singular with $1/f$ for an interpolation which is only transversely
continuous. This kind of improvement will have to await the
development of a smoother interpolation and the check of whether all the
local terms that might appear in $L(n)$ are locally integrable.

Finally, the gauge field interpolation as we have defined it is a rather
cumbersome procedure. A more economical one was given in  ref.\cite{inter}.
It satisfies all the criteria we give in
section 3,
with the possible exception of lattice rotational invariance, which we have
not been able to prove or disprove. However, this symmetry could be
restored by averaging over lattice rotations,
\begin{eqnarray}
\Gamma[U] \rightarrow \frac{\sum_{R} \Gamma[U^{R}]}{\sum_{R}},
\end{eqnarray}
where $U^R$ is a lattice rotation of $U$.
 It is important
to stress that cubic invariance is an essential
ingredient to recovering a Lorentz invariant continuum limit.

\section{Conclusion.}

In this paper, we have presented a non-perturbative lattice formulation of
 chiral gauge theories. It provides a finite algorithm for computer
simulations. The gauge fields live on a lattice with spacing $b$, taken as
the defining `cutoff' of the theory. For each gauge field configuration the
chiral fermion determinant is computed by first interpolating the gauge
field to a finer lattice with spacing $f \ll b$; the determinant is then
constructed in the background of this interpolated field. Though it is
difficult to maintain gauge invariance {\it and } eliminate the doubler
modes that plague lattice fermion constructions, the loss of gauge
invariance can be minimized using a separate
 `hybrid' regulator for the magnitude and phase of the fermion determinant.
The gauge non-invariance of the determinant is then of order $f^2$ so
long as gauge anomalies cancel.
The gauge field loops do not upset this suppression since their highest
momenta are of order $1/b$. There is one subtlety, that the gauge
field interpolation  is smooth enough to not reintroduce powers of
$1/f$ that can destroy the approximate gauge invariance. We checked that this
did not happen with the choice of `transversely continuous' interpolation
 made here. In the continuum
limit $f \rightarrow 0$, followed by $b \rightarrow 0$, exact gauge
invariance is recovered.

The price of
maintaining gauge invariance is the computational expense of gauge field
interpolation and calculation of large fermion determinants. We argued that
in order to be insensitive to the breakage of gauge invariance
 in a simulation in a physical volume $L^4$, one would have to
take
\begin{equation}
f/b \sim b/L.
\end{equation}
Strategies for improved efficiency were discussed including matching the
two-cutoff theory to a one-cutoff theory. The main computational effort
would then be in doing the matching. If this is possible,
 the one-cutoff theory could be
used for all amplitudes of interest, with the same cost as (unquenched)
 lattice QCD simulations!

We also  explored anomalies in global currents in our lattice construction
of chiral gauge theory, demonstrating the validity of the  Adler-Bardeen
theorem
to all orders. We briefly discussed the related physics of fermion
number violation, using $B+L$-violation in a simplified `standard model'
 as an example.

At present, the task of using the algorithm presented here to simulate
four-dimensional, non-abelian chiral gauge theories seems computationally
formidable. Of
course, we hope that there can be improvements made in the efficiency of the
algorithm, and in the ability of computers to carry them out. In the
meantime, simulation of two-dimensional chiral gauge theories seems quite
feasible and provides a useful arena for testing the method.

\section*{Acknowledgments}
We wish to thank S. Coleman, M. Golden, S. Hsu, S. Kachru,
J.P. Leroy, O. P\'ene,  C. Pittori and K. Rajagopal for discussions, and to
Ph. Boucaud and J.L. Alonso for a careful reading of our manuscript. We are
especially
grateful to C. Rebbi for discussions and encouragement. Finally, we want to
thank O. Narayan for repeatedly helping us in times of
mathematical  confusion.

\section*{Appendix A}

In this appendix we will prove in more technical detail the general bound on
one-fermion-loop lattice integrals (\ref{formfactor}), and the stricter
bounds (\ref{trans}) for the gauge invariant parity-even contributions
in $n=2,3$ and for the non-invariant parity-odd $n =3$ amplitudes.

These integrals have the form,
\begin{eqnarray}
\Gamma^{(n)}(q) = \int_{BZ(f)} d^4 k\; I(q,k,f) \equiv \int_{BZ(f)} d^4 k
\frac{V(q,k,f)}{\prod_i P^2_f(l_i(k,q),f)}
\end{eqnarray}
where the functions $P_f$ are the lattice fermion propagators, so
\begin{eqnarray}
P^2_f(k) \equiv \frac{1}{\sum_\mu  {\hat k}_\mu^2  + M(k)^2}, \nonumber\\
M(k) \equiv \frac{2 r}{f} \sum_\mu sin^2(k_\mu f/2) \;\;\; {\hat k}_\mu \equiv
\frac{1}{f} sin k_\mu f.
\end{eqnarray}
The fermion momentum is $k \in BZ(f) \equiv (\frac{-\pi}{f},\frac{\pi}{f})$.

It has been proven in \cite{luscher2} for very general conditions on the
 functions
$V$ and $P$, which our action satisfies, that the integrand can be bounded
by a continuous function that does not depend on $f$, for all $f < f_0$ and
$f_0$ finite and fixed. We define the integer $\omega$,
\begin{eqnarray}
V(q,k,f) = f^{-\omega} {\tilde V}(q f, kf)
\end{eqnarray}
and the degree of divergence of V as,
\begin{eqnarray}
lim_{\lambda\rightarrow \infty} \;\; V(q, k \lambda, f/\lambda) \; =  K
\lambda^\nu + O(\lambda^{\nu-1})
\end{eqnarray}
and $K \neq 0$. For the integrals we consider here, the following equality
holds,
\begin{eqnarray}
2 I_f -\omega  = n
\end{eqnarray}
with $I_f$ the number of internal fermion lines and $n$ being the number
of external boson lines.
The last equality can be easily checked from the lattice Feynman rules for
fermion loop integrals \cite{rothe}.

With these definitions,  L\"uscher's result is
\begin{eqnarray}
|V(q,k,f)| < C |q|^{(\omega-\nu)} Q^\nu(|k|,|q|),
\end{eqnarray}
where $Q$ is a homogeneous polynomial of degree $\nu$ and $C$ is a constant
independent of $f$.
On the other hand, for each propagator we get,
\begin{eqnarray}
|P^2_f(l,f)| > C' l^2,
\end{eqnarray}
where $C'$ is some constant and $l^2$ is the square of the fermion momentum.

Then,
\begin{eqnarray}
|\Gamma^{(n)}(q)| < \int_{BZ(f)} d^4 k \frac{C |q|^{(\omega -\nu)}
Q^\nu(|k|,|q|)}{C' \prod_i l_i(q,k)^2}.
\label{boundint}
\end{eqnarray}
If $I_f$ is the number of internal fermion lines, we can distinguish
two possibilities. If $d_{uv} < 0$,
\begin{eqnarray}
d_{uv} \equiv 4 + \nu - 2 I_f < 4 - n
\end{eqnarray}
the integral is convergent in the limit $f \rightarrow 0$ and by power counting
for continuum integrals this implies,
\begin{eqnarray}
|\Gamma^{(n)}| < |q|^{4+\omega-2 I_f} Log(|q|) = |q|^{4-n} Log(|q|),
\end{eqnarray}
where $n$ is the number of external boson lines.

In the case $d_{uv} \geq 0$, the integral (\ref{boundint}) is
divergent in the limit $f\rightarrow 0$, so it must depend on the cutoff
$1/f$.
Note however, that we can separate the divergent parts from the
finite part by an expansion at zero external momenta. From (\ref{boundint})
we see that there are no infrared problems if
\begin{eqnarray}
d_{ir} \equiv 2 I_f - 4 -\omega < 0,
\end{eqnarray}
but $d_{ir} = - d_{uv} + \nu -\omega \leq -d_{uv}$. Since the degree of
infrared divergence increases with each derivative with respect to external
momenta, while the ultraviolet $d_{uv}$ decreases, we see that for
$uv$ divergent
diagrams, we can Taylor expand up to $d_{uv} -1$ times or more. The local
terms in the expansion are substracted and the remainder is only
logarithmically divergent, and also possibly non-local. In our hybrid
scheme and if the
gauge anomalies cancel, there is no
local counterterm except the one contained in this remainder, which is
logarithmic and cannot be substracted at zero momentum.
Only three cases need to be considered $n=2,3,4$. In the case of
$n=4$, $d_{uv} \leq 0$ so the integral (\ref{boundint}) is at
most logarithmically
dependent on the cutoff,
\begin{eqnarray}
|\Gamma_0^{(n)}| < C Log(|q| f).
\end{eqnarray}
We can easily renormalize it by substracting the same integral at zero
 external momenta, with an infrared cutoff $\sim 1/b$,
\begin{eqnarray}
\delta \Gamma^{(4)}(0) \equiv \int_{BZ(b)->BZ(f)} d^4 k \; I(0,k,f).
\end{eqnarray}
(This corresponds to the gauge coupling renormalization of eq.
(\ref{coupling}).)
The remainder is then finite. This is easy to prove using the L\"uscher
bounds. The renormalized integral is then,
\begin{eqnarray}
\Gamma^{(4)}(q) = \int_{BZ(b)} d^4 k I(q,k,f)
+ \int_{BZ(f)-BZ(b)} d^4 k (I(q,k,f)-I(0,k,f))
\label{renor}
\end{eqnarray}
The integrand of the second term obviously vanishes for zero external momenta,
so it can be written as a Taylor remainder of a first order expansion around
zero external momenta. This remainder is of the form,
\begin{eqnarray}
R_1 = \sum_{\alpha} q_\alpha \int^1_0 dt (1-t) \frac{\partial}{\partial
w_\alpha} I(q+tw,k,f) |_{\omega =0},
\end{eqnarray}
and now $R_1$ has $d_{uv} = -1$, so the second term is finite in the limit
$f\rightarrow 0$, by (\ref{boundint}). The first term obviously is too.
{}From (\ref{renor}) we then find,
\begin{eqnarray}
|\Gamma^{(4)}| < C Log(|q| b),
\end{eqnarray}
which is finite in the limit $f\rightarrow 0$ and has a logarithmic bound
for large external momenta as expected.

The cases of $n=2$ and $n=3$, since they are only logarithmically divergent
for the hybrid action, can be analyzed in the same way and the following
bounds are found,
\begin{eqnarray}
|\Gamma^{(2)}| < C |q|^2 Log(|q| b); \;\;\;\; |\Gamma^{(3)}| < C |q| Log(|q|
b).
\label{g23a}
\end{eqnarray}

However, as we saw these bounds are not good enough to prove $f$ convergence
due to the discontinuities of the interpolation, and the stricter bounds
(\ref{trans}) are needed. In the continuum, we know that gauge invariance
implies that the
two-point function is of the form,
\begin{eqnarray}
\Gamma^{(2)}_{\mu\nu}(q) = (q_\mu q_\nu - g_{\mu\nu} q^2) \Pi(q)
\end{eqnarray}
where $\Pi(q)$ is a scalar function that behaves logarithmically for large
external momenta. It is clear then that the component
$\mu=\nu$ satisfies the bound (\ref{trans}).
For finite $f$ however, a form like this is not easy to prove for the
parity-even parts, because
we do not have Lorentz invariance.
Fortunately, the gauge invariant relations of eqs. (\ref{gipe}) are enough.
Let us start with the two-point function. Using (\ref{gipe}) repeatedly we
find,
\begin{eqnarray}
\Gamma^{(2)}_{\mu\mu}(q) = \hat{q}^2 \frac{\Gamma^{(2)}_{\mu\mu}(q)}{\hat{q}^2}
= \sum_{\beta\neq\mu} {\hat q}^2_\beta
\frac{\Gamma^{(2)}_{\mu\mu}}{{\hat q}^2} -
\sum_{\alpha,\beta\neq\mu} \hat{q}_\alpha \hat{q}_\beta
\frac{\Gamma^{(2)}_{\alpha\beta}}{\hat{q}^2}
\end{eqnarray}
where the usual definition,
\begin{eqnarray}
\hat{q}_\alpha \equiv \frac{e^{i q_\alpha f} -1}{f},
\end{eqnarray}
has been used. Now, the right-hand-side can easily be bounded and we find
the desired result,
\begin{eqnarray}
|\Gamma^{(2)}_{\mu\mu}(q)| < \sum_{\alpha,\beta} |q_\alpha| |q_\beta|
|\Gamma^{(2)}_{\alpha\beta}| \frac{1}{q^2} < \sum_{\alpha,\beta} |q_{\alpha}|
|q_{\beta}| Log(|q|),\;\; \;\; \alpha,\beta \neq \mu.
\end{eqnarray}
The last inequality has made use of eq. (\ref{g23a}). The P-even
three-point function can be analysed in the same way.

Finally, the parity-odd three point function has to be considered. In this
case we do not have the help of the gauge invariance constraints (\ref{gipe}),
but on the other hand, the diagram contains an epsilon tensor.
A direct inspection of the three-point function reveals the following
structure,
\begin{eqnarray}
\Gamma^{(3)}_{\mu\mu\mu}(q,q') \sim f^{abc} \epsilon_{\mu\alpha\beta\gamma}
\{ 4 \bar{s}(q'_\beta) \bar{s}(q_\gamma)
F_1(q,q') - 2 \bar{s}(q_\gamma)
s(q'_\beta) F_2(q,q') \nonumber\\
+ 2 s(q_\gamma)  \bar{s}(q'_\beta) F_3(q,q')
-   s(q_\gamma) s(q'_\beta) F_4(q,q') \},
\end{eqnarray}
where
\begin{eqnarray}
\bar{s}(q) \equiv \frac{sin^2(q f/2)}{f^2} \;\;
s(q) \equiv \frac{sin(q f)}{f},
\end{eqnarray}
and the functions $F$ are lattice integrals with smaller degree of divergence
than the original integral, for which
the following bounds hold,
\begin{eqnarray}
|F_1(q,q')| <  \frac{1}{|q|^3} Log(|q|), \; |F_{2,3}(q,q')| <
\frac{1}{|q|^2} Log(|q|)\nonumber\\ |F_4(q,q')| < \frac{1}{|q|} Log(|q|).
\end{eqnarray}
Using $|\bar{s}(q)| < C |q|^2$, $|s(q)| < C |q|$,
it is then straightforward
to check that
\begin{eqnarray}
|\Gamma^{(3)}_{\mu\mu\mu}(q,q')| < |q_\alpha| Log(|q|), \;\;\; \alpha \neq \mu,
\end{eqnarray}
This concludes the proof of the bound (\ref{gipe}) for the P-odd
three-point function.


\begin{thebibliography}{99}
\bibitem{rev} For examples of other  recent activity see   the articles
appearing in Nucl.Phys. B(Proc.Suppl)
{\bf 34}(1994).
\bibitem{early} R. Flume and D. Wyler, Phys. Lett.{\bf 108B}(1982)317.
\bibitem{thooft} G. 't Hooft, Phys.Lett.{\bf B349}(1995)491.
\bibitem{schier} M. Gockeler, G. Schierholz, Nucl.Phys. B(Proc.Suppl)
{\bf 29B,C}(1992) 114; Nucl.Phys. B(Proc.Suppl)
{\bf 30}(1993) 609.
\bibitem{hsu} S. Hsu, hep-th/9503058.
\bibitem{kron} A. Kronfeld, FERMILAB-PUB-95/073-T or hep-lat/9504007.
\bibitem{nielsen} H.B. Nielsen and H. Ninomiya, Nucl. Phys.
{\bf B185}(1981)20,(E) {\bf B195}(1982) 541 and Nucl. Phys. {\bf
B193}(1981)173.
\bibitem{smit} L.H. Karsten and J. Smit. Nucl. Phys. {\bf B183}(1981) 103-140.
\bibitem{rome} A. Borrelli, L. Maiani, G.C. Rossi, R. Sisto and M. Testa,
 Nucl. Phys. {\bf B333}(1990) 335-356.
\bibitem{rome2} L. Maiani, G.C. Rossi and M. Testa, Phys. Lett. {\bf
B292}(1992) 397.
\bibitem{zaragoza} J.L. Alonso, Ph. Boucaud, J.L. Cortes and E. Rivas, Mod.
Phys. Lett. {\bf A5}(1990) 275; Nucl.Phys. B(Proc.Suppl)
{\bf 17}(1990) 461.
\bibitem{zumino} J. Wess and B. Zumino, Phys. Lett. {\bf 37B}(1971) 95;
 W.A. Bardeen and B. Zumino, Nucl. Phys. {\bf B244}(1984) 421-453.
\bibitem{rothe} See for instance Lattice Gauge Theories by H.J. Rothe, World
Scientific Lecture Notes in Physics-Vol {\bf 43} 1992.
\bibitem{golt} M. Golterman and D. Petcher, Phys. Lett. {\bf 225}(1989) 159.
\bibitem{gaume} L. Alvarez-Gaum\'e and S. Della Pietra, in {\it Recent
Developments in Quantum Field Theory}, eds. J. Ambjorn, B.J. Durhuus and
J.L. Petersen, North-Holland (1985).
\bibitem{bodwin} G.T. Bodwin and E.V. Kovacs, Nucl.Phys. B(Proc.Suppl)
{\bf 30}(1993) 617.
\bibitem{witten} E. Witten, Phys. Lett. {\bf 117B}(1982) 324.
\bibitem{luscher1} M. L\"uscher, Comm. Math. {\bf 85}, 39-48 (1982).
\bibitem{reisz} T. Reisz, Comm. Math. {\bf 117}, (1988a) 79; {\bf 117}(1988b)
639;
{\bf 116} (1988c) 81.
\bibitem{luscher2} M. L\"uscher, Lectures at Summer School on Fields,
Strings and Critical Phenomena, Les Houches (1988).
\bibitem{smit3} J. Smit, Nucl.Phys. B(Proc.Suppl)
{\bf 29B,C}(1992) 83.
\bibitem{adler} S. Adler and W.A. Bardeen, Phys. Rev. {\bf 182} (1969) 1517.
\bibitem{kawai} H. Kawai, R. Nakayama and K. Seo, Nucl. Phys. {\bf B189},
40 (1981).
\bibitem{overlap} R. Narayanan and H. Neuberger, Nucl. Phys. {\bf B443}(1995)
305.
\bibitem{thooft2} G. 't Hooft, Phys. Rev. Lett. {\bf 37} (1976) 8.
\bibitem{ringwald} V. Zakharov, Nucl. Phys. {\bf B371} (1992) 637-658.
M. Porrati, Nucl. Phys. {\bf B347}(1990) 371-393. V.V. Khoze and
A. Ringwald, Nucl. Phys. {\bf B355}(1991) 351.
\bibitem{fuji} K. Fujikawa, Phys. Rev {\bf D21} (1980) 2848 and {\bf 29} (1984)
285.
\bibitem{atiyah} M.F. Atiyah and I.M. Singer, Ann. Math. {\bf 87}, 484 (1968).
\bibitem{inter} M. G\"ockeler, A. Kronfeld, G. Schierholz and U.J.Wiese,
Nucl. Phys. {\bf B404}(1993) 839.
\end{thebibliography}
\end{document}